\documentclass[acmsmall]{acmart}

\usepackage{xcolor}

\AtBeginDocument{%
  }

\setcopyright{cc}
\setcctype{by}
\acmJournal{PACMMOD}
\acmYear{2025} \acmVolume{3} \acmNumber{6 (SIGMOD)} \acmArticle{290} \acmMonth{12} \acmPrice{}\acmDOI{10.1145/3769755}

\acmDOI{10.1145/3769755}

\usepackage{hyperref}
\usepackage[ruled,linesnumbered]{algorithm2e}
\usepackage{enumitem}
\usepackage{dsfont}
\usepackage{dutchcal}
\usepackage{cancel}
\usepackage[normalem]{ulem}
\usepackage{makecell}
\usepackage{multirow}
\usepackage{amsmath}
\usepackage{pict2e}
\usepackage{subcaption}
\usepackage{multicol}

\def\header{\vspace{1mm} \noindent}

\newcommand{\figref}[1]{Fig.\ \ref{#1}}
\newcommand{\tabref}[1]{Table~\ref{#1}}
\newcommand{\secref}[1]{Sec.\ \ref{#1}}
\newcommand{\algref}[1]{Alg.\ \ref{#1}}

\newcommand{\proposed}{ACGraph}

\newcommand{\degt}{\delta_{deg}}
\newcommand{\idt}{\theta_{id}}

\newcommand{\revise}[1]{{#1}}

\newcommand{\reviseROne}[1]{{#1}}
\newcommand{\reviseRTwo}[1]{{#1}}
\newcommand{\reviseRThree}[1]{{#1}}

\begin{document}

\title{ACGraph: An Efficient Asynchronous Out-of-Core Graph Processing Framework}

\author{Dechuang Chen}
\affiliation{%
  \institution{The Chinese University of Hong Kong}
  \country{Hong Kong, China}
}
\email{dcchen@se.cuhk.edu.hk}

\author{Sibo Wang}\authornote{Sibo Wang and Qintian Guo are the corresponding authors.}
\affiliation{%
  \institution{The Chinese University of Hong Kong}
  \country{Hong Kong, China}
}
\email{swang@se.cuhk.edu.hk}

\author{Qintian Guo}\authornotemark[1]
\affiliation{%
  \institution{The Hong Kong University of Science and Technology}
  \country{Hong Kong, China}
}
\email{qtguo@ust.hk}

\begin{abstract}

Graphs are a ubiquitous data structure in diverse domains such as machine learning, social networks, and data mining. As real-world graphs continue to grow beyond the memory capacity of single machines, out-of-core graph processing systems have emerged as a viable solution. Yet, existing systems that rely on strictly synchronous, iteration-by-iteration execution incur significant overheads. In particular, their scheduling mechanisms lead to I/O inefficiencies, stemming from read and work amplification, and induce costly synchronization stalls hindering sustained disk utilization. To overcome these limitations, we present {\em ACGraph}, a novel asynchronous graph processing system optimized for SSD-based environments with constrained memory resources. ACGraph employs a dynamic, block-centric priority scheduler that adjusts in real time based on workload, along with an online asynchronous worklist that minimizes redundant disk accesses by efficiently reusing active blocks in memory. Moreover, ACGraph unifies asynchronous I/O with computation in a pipelined execution model that maintains sustained I/O activation, and leverages a highly optimized hybrid storage format to expedite access to low-degree vertices. We implement popular graph algorithms, such as Breadth-First Search (BFS), Weakly Connected Components (WCC), personalized PageRank (PPR), PageRank (PR), and $k$-core on ACGraph and demonstrate that ACGraph substantially outperforms state-of-the-art out-of-core graph processing systems in both runtime and I/O efficiency. 

\end{abstract}

\received{April 2025}
\received[revised]{July 2025}
\received[accepted]{August 2025}





\maketitle

\section{Introduction} \label{sec:intro}

Graphs are a powerful data structure widely used in machine learning \cite{kipf2017semi,hamilton2017inductive,velickovic2018graph}, social networks \cite{newman2003structure,watts1998collective}, data mining \cite{leskovec2007graph,gspan}, among other domains. The need for fast processing of large-scale graphs has become a critical requirement, giving rise to a series of parallel graph processing systems (GPS) \cite{galois, ligra, corograph, pregel, gpop}. Yet, as real-world graphs, such as social networks, web graphs, e-commerce networks, and recommendation graphs, continue to grow in size and complexity, they often exceed the memory capacity of a single machine. This challenge has led to the development of both distributed graph computing frameworks \cite{powergraph,graphx,gemini} and out-of-core GPSs \cite{graphchi,flashgraph,xstream,GridGraph,Blaze,CAVE,clip}. Although distributed frameworks offer strong scalability, the substantial overhead from fault tolerance, network communication, and load balancing limits their overall performance potential. By contrast, single-machine out-of-core GPSs achieve a favorable trade-off between scalability and efficiency. Prior work has demonstrated that these systems deliver higher performance per unit cost, making them an attractive alternative for many applications \cite{GridGraph,graphchi,flashgraph,graphene, mosaic}.

\reviseRTwo{
Early out-of-core GPSs, such as GraphChi \cite{graphchi}, X-Stream \cite{xstream}, and GridGraph \cite{GridGraph}, were specifically designed for hard disk drives (HDDs). Due to the poor random access performance and limited bandwidth of HDDs, optimizing access locality of edges and vertices becomes critical for system performance.  To achieve this, they often sacrifice computational efficiency and incur redundant I/O overhead, prioritizing sequential disk access. For instance, GridGraph partitions graphs into 2-D grids and processes each grid entirely in memory to maximize sequential disk access, but at the cost of significant additional disk reads.
}

Nowadays, advancements in storage technology and declining hardware costs have enabled the adoption of SSDs for large-scale graph storage and processing. Modern SSDs provide orders-of-magnitude faster read/write speeds and near-uniform random/se-quential access performance. \reviseRTwo{Unfortunately, these HDD-era frameworks have proven unable to directly leverage the advantages of these faster storage devices, without substantial redesign \cite{mosaic}.} Recent {SDD-optimized} out-of-core GPSs focus optimization efforts on minimizing overall I/O volume while fully leveraging the available SSD bandwidth. For instance, CAVE \cite{CAVE} leverages intra-/inter-subgraph parallelization approaches to fully exploit the capabilities of underlying storage devices, whereas Blaze \cite{Blaze} employs a pipelined architecture with a new scatter-gather technique called online binning to overlap computation with I/O operations, ensuring uninterrupted data access within an iteration.
\reviseRTwo{Experimental results demonstrate that these SDD-era designs consistently outperform those HDD-era designs on high-performance SSDs \cite{flashgraph, mosaic, CAVE}.}

However, these GPSs rely on strict synchronization semantics that enforce iteration-by-iteration execution, leading to two key limitations. The first is \textit{I/O inflation}, where the actual amount of data retrieved via I/O operations far exceeds what the task initially requires. This phenomenon primarily stems from two sources: \textit{read inflation} and \textit{work inflation}. Read inflation occurs when accessing a small amount of data (e.g., a few neighboring vertices) requires transferring an entire SSD block. More critically, synchronization semantics prevent accesses to the same disk block across different iterations from being merged into a single I/O operation. Furthermore, these accesses typically have long reuse intervals, as they are interleaved with numerous accesses to other blocks. This long-interval access pattern reduces temporal locality and wastes fetched data. The work inflation arises when a synchronous algorithm incurs higher workload than its sequential counterpart, often because the sequential one leverages prioritized vertex processing to complete the task more efficiently. We empirically validate both sources of overhead in synchronous GPSs in \secref{sec:overhead1}.

The second limitation lies in synchronization stalls imposed by strict  barrier requirements. In these systems, a new iteration cannot begin until the entire active set of vertices (i.e., the processing frontiers) has been fully processed. The system must pause between iterations to wait for all worker threads to complete their tasks, aggregate the newly activated vertices for the next iteration, and then redistribute them to balance the load. Due to memory constraints, I/O threads remain idle until worker threads have processed loaded data blocks and freed up memory. This bottleneck impedes sustained disk utilization. As will be shown in \secref{sec:overhead2}, synchronous systems like Blaze \cite{Blaze} and CAVE \cite{CAVE} show intermittent low disk activity between iterations of BFS, while our proposed ACGraph achieves persistent disk saturation after a short initialization phase.

\header
{\bf Contribution.} Prior theoretical analyses of graph algorithms have shown that many graph algorithms can be executed in both synchronous and asynchronous modes \cite{CormenG96,parllelcomputing,XieCGZC15}. \reviseRThree{This motivates us to consider designing efficient out-of-core asynchronous GPSs. However, directly porting existing in-memory asynchronous paradigm to out-of-core environments exposes fundamental challenges. Out-of-core execution makes data movement a first-order cost, so existing asynchronous schedulers, designed for memory-resident data, fails: coordinating fine-grained task execution with preloading and eviction requires rethinking scheduling. Moreover, ignoring access locality leads to severe SSD read inflation and under-utilization of fetched blocks, as vertices sharing blocks become scattered across priority queues. At the same time, in-memory asynchronous optimizations (e.g., fine-grained parallelism) conflict with out-of-core efficiency needs (e.g., batched I/O), and higher disk latency demands co-design of task scheduling and data movement to overlap computation with I/O. } Motivated by these insights, we propose \textbf{ACGraph} (\underline{A}synchronous Out-of-\underline{C}ore \underline{Graph} Processing System), a novel asynchronous GPS optimized for SSDs with constrained memory. ACGraph uses the \textit{semi-external} model, where vertex data fits in memory while the full edge set does not and resides on disk, which has been widely adopted in prior work \cite{CAVE, Blaze, chunkgraph, flashgraph, mosaic, grafu, graphene, ZhuXWL13}.

Unlike conventional frameworks that rely on fixed or coarse-grained scheduling, ACGraph introduces a dynamic, block-centric priority scheduler that adjusts block priorities in real time based on workload dynamics. ACGraph distinguishes itself through an online asynchronous worklist that continuously monitors and reuses loaded blocks immediately upon reactivation, thereby reducing redundant disk accesses and adapting dynamically to changing graph workloads. Further, by unifying asynchronous I/O with computation into a seamless pipelined execution model, our {ACGraph} achieves continuous execution and high I/O throughput. Finally, a highly optimized hybrid graph storage format is dedicated to ACGraph that streamlines access to low-degree vertices without additional memory or computational overhead. Collectively, these innovations enable ACGraph to deliver significant runtime and I/O improvement compared to state-of-the-art systems under the memory constraints of modern SSD-based platforms.

We implement multiple popular algorithms on ACGraph, including Breadth-First Search (BFS), Weakly Connected Component (WCC), personalized PageRank (PPR), PageRank (PR)\footnote{Note that PR is a special case of PPR with a uniform initial distribution and can be derived using PPR algorithms \cite{page1999pagerank, FORA, 0001YWXWLY019}.}, and $k$-core, and compare it against two recent out-of-core GPSs, CAVE \cite{CAVE} and Blaze \cite{Blaze}. Experimental results show that ACGraph significantly outperforms both. Our contributions are as follows:
\begin{itemize}[leftmargin=*]
    \item We systematically identify that current out-of-core GPS struggles to achieve high I/O efficiency due to I/O inflation and synchronization stalls caused by the synchronization semantics.
    \item We propose ACGraph, a novel asynchronous out-of-core GPS with block-based priority scheduling, which eliminates synchronization overhead. A novel worklist is designed to dynamically reuse loaded blocks, reducing redundant disk I/O while adapting to evolving graph workloads.
    \item We propose a highly optimized hybrid storage architecture that enhances access efficiency for low-degree vertices, while simultaneously reducing the graph storage cost, complemented by targeted optimizations to ensure computational efficiency and reduced memory overhead.
    \item We evaluate ACGraph against state-of-the-art out-of-core GPSs on various workloads including BFS, WCC, $k$-core, PPR, and PR, and show that ACGraph significantly improves over competitors. Our ACGraph achieves up to 12$\times$ speedup over CAVE and 15$\times$ speedup over Blaze under similar or even lower memory costs.
\end{itemize}

\section{Preliminary}\label{sec:background}

Consider a directed graph $G = (V, E)$, where $V$ is the vertex set and $E$ is the edge set. An edge from vertex $u$ to vertex $v$ is denoted as $(u, v)$. Without loss of generality, we focus on directed graphs, as undirected graphs can be transformed into a directed one by replacing each edge with two directed ones, an approach commonly used in graph processing \cite{forkgraph, DBLP:journals/csur/ShiZZJHLH18}. Let $N(v)$ denote the set of out-neighbors of vertex $v$, and $deg(v) = |N(v)|$ its out-degree. 

\subsection{Graph Processing System (GPS)}
\begin{figure}[t]
    \centering
    \hspace{-2mm}
    \includegraphics[height=29mm]{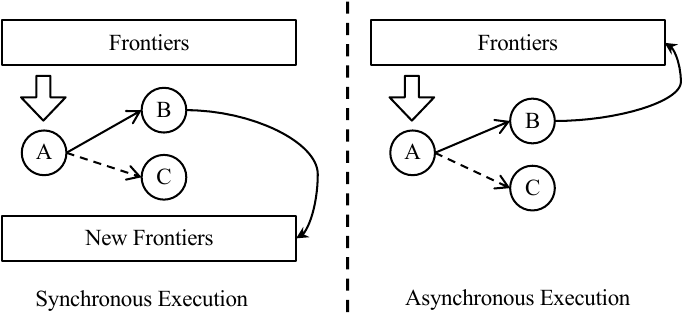}
    \vspace{-1mm}
    \caption{The vertex-centric synchronous/asynchronous execution model, where vertex $A$ is being processed and passes message to its neighbors $B$ and $C$, while only $B$ gets activated.}
    \label{fig:vcent}
    \vspace{-3mm}
\end{figure}

Most GPSs are built on the consensus that many graph algorithms follow an iterative message passing pattern: Each iteration starts with a set of active vertices (the \textit{frontier}) that explore their out-neighbors. Neighboring vertex states are updated, and those meeting activation criteria form the new frontier. This process repeats until no new vertices are activated, and the algorithm \textit{converges}.

As an example, we apply the Label Propagation algorithm to compute WCC. Each vertex maintains a WCC label, initially set to its own ID. 
All vertices start in the initial frontier. 
In each iteration, a frontier vertex $v$ propagates its label to out-neighbors, who compare it with their own. If the received label is smaller, the out-neighbor updates its label, becomes \textit{activated}, and joins the next frontier. The process continues until there are no new frontiers.

The above WCC example naturally motivates the \textit{vertex-centric} execution model, where vertices serve as the fundamental scheduling units. 
As illustrated in \figref{fig:vcent}, a worker thread is processing vertex $A$ that belongs to the current frontier set. After processing, vertex $B$ is activated (indicated by a solid line) and is immediately inserted into the frontier set for the next iteration, while vertex $C$ remains inactive (as shown by a dashed line).
In the rest of this paper, the terms \textit{active vertex} and \textit{frontier} are used interchangeably to denote vertices requiring processing.

\header
\textbf{Synchronous/Asynchronous GPSs.} The execution models of GPS can be categorized as either synchronous or asynchronous, as illustrated in \figref{fig:vcent}. In a synchronous execution model, explicit barriers are inserted between iterations to ensure that all frontiers are completely processed before the next iteration begins. For example, as shown in the left side of \figref{fig:vcent}, the new frontier set are generated based on the outcome of the fully processed frontiers from the preceding iteration. In contrast, asynchronous execution model removes these strict barriers and often incorporates priority scheduling to improve efficiency for certain algorithms. For instance, as shown in the right side of \figref{fig:vcent}, once vertex $B$ is activated, it is immediately inserted into the frontier set, typically maintained as a list. Note that vertex $B$ may be scheduled before vertices from prior iterations, if it has high priority. To illustrate further, when executing the WCC algorithm, prioritizing vertices with the smallest label usually lead to faster convergence.  

Most existing out-of-core GPSs adopt a synchronous model to enable pre-identification of required disk blocks and coordinated I/O-computation scheduling. However, this convenience comes at the cost of significant I/O inefficiency and synchronization overhead, as detailed in \secref{sec:motivation}.

\header\textbf{Asynchronous I/O.} Modern NAND-based SSDs use multi-layer architectures and parallel access across multiple flash chips to boost data transfer bandwidth \cite{DBLP:conf/fast/JunPKKS24, DBLP:journals/jsa/KangKPPL07, DBLP:conf/usenix/AgrawalPWDMP08}. They typically use 4KB as the minimum I/O unit and offer comparable performance for sequential and 4KB random reads \cite{DBLP:journals/pvldb/HaasL23}. Leveraging this, our ACGraph adopts 4KB blocks as the basic partitioning unit for graph data.

Synchronous I/O incurs performance bottlenecks, as threads block on I/O, causing frequent context switches and underutilized CPU resources. Fortunately, asynchronous I/O, such as \texttt{io\_uring} \cite{iouring}, addresses this by using dual ring buffers shared between user and kernel space, enabling zero-copy transfer and reducing syscall overhead. Our ACGraph integrates \texttt{io\_uring} asynchronous mechanism, successfully overlapping computational tasks and I/O operations to establish an efficient pipeline execution architecture.

\subsection{Existing Out-of-Core GPSs} \label{sec:oocgps}
Out-of-core GPSs have been widely studied \cite{graphchi, xstream, GridGraph, flashgraph, mosaic, CAVE, Blaze, clip, lumons, grafu, seraph, minigraph}. Earlier systems \cite{graphchi,GridGraph,xstream}, like GridGraph \cite{GridGraph}, target HDDs and exploit cheap sequential scans, often at the cost of extra I/O, to avoid random accesses. Later work focuses on reducing I/O costs. \reviseRThree{Some systems accelerate convergence by asynchronous semantics \cite{clip, wonderland, asyncstripe}. For example, CLIP~\cite{clip} maximizes I/O efficiency by reusing loaded data across multiple processing steps. In contrast, Grafu \cite{grafu} reduces I/O under synchronous semantics, achieving up to 50\% lower runtime and I/O costs both theoretically and in practice compared to GridGraph. However, all of these methods rely on GridGraph-style partitioning, using large grid partitions as the basic scheduling unit, which leads to additional I/O overhead.} \reviseROne{Besides, MiniGraph \cite{minigraph} adopts a hybrid vertex-/graph-centric model with pipelined I/O and computation to boost parallelism. Yet, like GPSs mentioned above, it partitions the graph into memory-resident fragments as scheduling units and issues a few bulk-load I/Os to reduce random-access, still incurring unnecessary I/Os. As noted in \secref{sec:intro}, such design require substantial redesign for SSDs.}

The superior bandwidth and random access performance of modern SSDs have shifted the focus to maximizing device utilization and minimizing I/O operations \cite{graphene, flashgraph, mosaic, CAVE, Blaze, seraph}. Graphene \cite{graphene} introduces four techniques: an I/O-centric programming model, bitmap-based asynchronous I/O, direct hugepage support, and workload balancing, to boost I/O performance. FlashGraph~\cite{flashgraph} overlaps computation with I/O and optimizes bandwidth through selective access, compact external-memory data structures, SSD access rescheduling to improve page cache hit rates, and conservative I/O merging to reduce CPU overhead. Mosaic \cite{mosaic} is a new locality-optimizing, space-efficient graph representation called Hilbert-ordered tiles, and a hybrid execution model that enables vertex-centric operations in fast host processors and edge-centric operations in massively parallel coprocessors (e.g. Xeon Phi). \reviseRThree{These frameworks adopt the semi-external setting, consistent with the setting of {\proposed}. In contrast, Seraph \cite{seraph} targets on-demand processing in a \textit{fully-external} setting where even vertex data does not fit in memory, unlike our semi-external approach.} Blaze \cite{Blaze} and CAVE \cite{CAVE} are recent out-of-core GPSs under the semi-external model, specifically optimized for SSDs. As shown in their evaluations, both systems deliver superior performance compared to alternative solutions. Thus we take them as our main competitors.

\section{Key Observations}\label{sec:motivation}

As discussed in \secref{sec:intro}, we identify two primary limitations that hinder the development of out-of-core processing systems under synchronous semantic models: I/O inflation and synchronization stalls. This section elaborates on these challenges in detail, which serves as the key motivation for our ACGraph framework.

\subsection{I/O Inflation} \label{sec:overhead1}
We use the term \textbf{I/O inflation} to describe the situation where the actual amount of data retrieved through I/O operations is significantly greater than the amount of data originally requested by the application or task.
The I/O inflation under synchronous execution primarily stems from two sources: \textbf{read inflation} and \textbf{work inflation}, which we describe in detail shortly. 

To empirically investigate these two factors and evaluate why conventional buffer pool technique fail to mitigate them, we design the following experiment: we run BFS and WCC algorithms under synchronous execution modes and record disk block accesses at each iteration. Having full knowledge of the block access patterns, we implement various cache replacement strategies to reduce the total number of I/O accesses, as follows:
\begin{enumerate}[leftmargin=*, topsep=3pt]
    \item Belady's optimal algorithm (\textit{OPT}) \cite{belady, osconcept}: A theoretically optimal policy that evicts the item with the longest time until its next access, establishing a minimal disk access baseline.
    \item A suboptimal policy (\textit{SUB}): A heuristic that evicts blocks unused in the next iteration (when identifiable) or randomly selects victims otherwise.
    \item Least Recently Used (\textit{LRU}): A widely adopted policy that prioritizes eviction of the least recently accessed items.
\end{enumerate}

\begin{figure}[!t]
\begin{small}
    \begin{tabular}{cccc}
    \multicolumn{4}{c}{\includegraphics[width=0.6\linewidth]{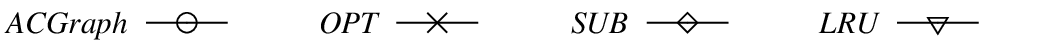}}\\
    \hspace{-2mm}
    \includegraphics[height=22mm]{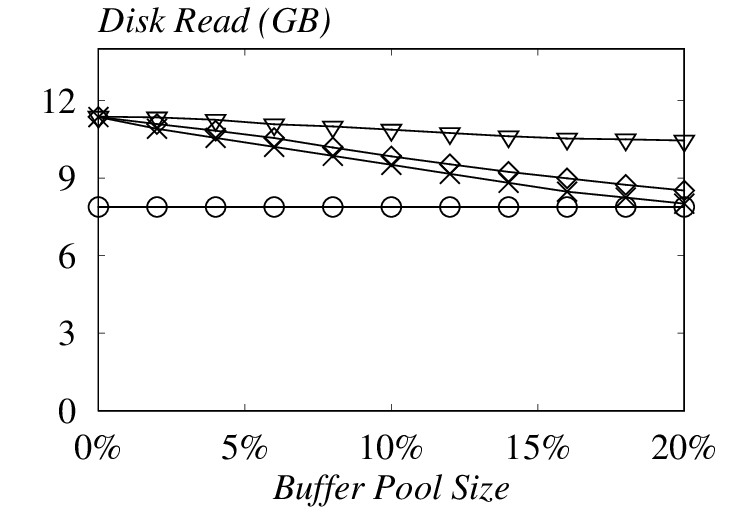} &
    \hspace{-2mm}
    \includegraphics[height=22mm]{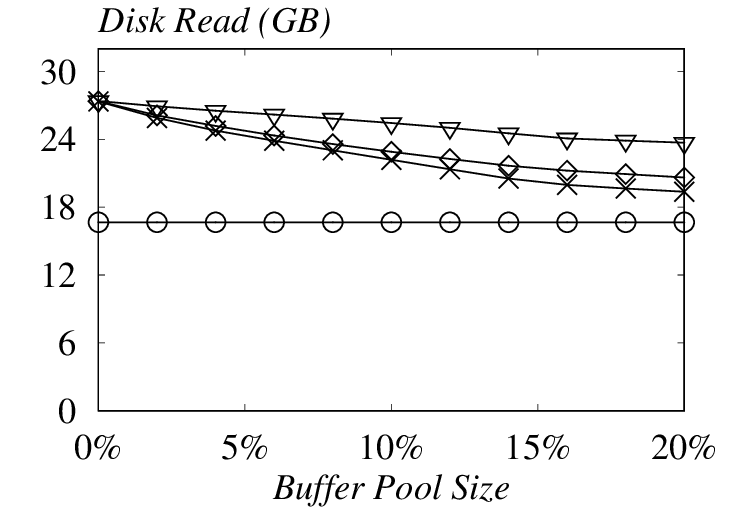} &
    \hspace{-2mm} 
    \includegraphics[height=22mm]{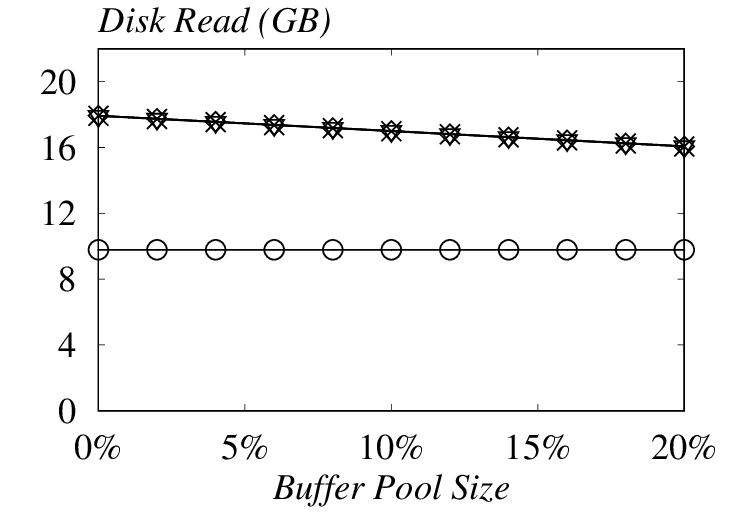} &
    \hspace{-2mm}
    \includegraphics[height=22mm]{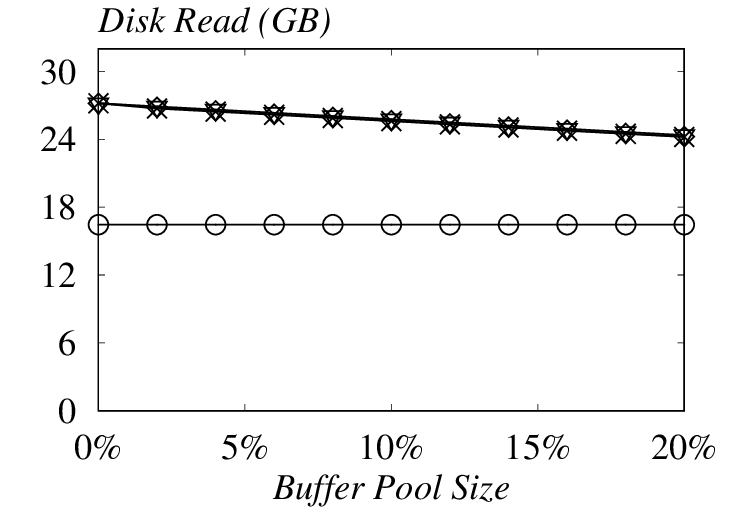}	\\ [-1mm]
    (a) Twitter (BFS) & (b) Friendster (BFS) & (c) Twitter (WCC) & (d) Friendster (WCC)  \\
    \end{tabular}
    \vspace{-1mm}
    \caption{Disk read volume vs. buffer size ($\%$ of graph size).}
    \label{fig:overhead1}
    \vspace{-4mm}
\end{small}
\end{figure}

As the \textit{OPT} policy assumes full future access knowledge (albeit unrealistic), it serves as a theoretical lower bound for the disk read volume under synchronous execution. \figref{fig:overhead1} shows the total disk read volume under the three cache replacement strategies, as buffer pool size varies from 4 MB to $20\%$ of the graph size (with minimum size 4 MB denoted as 0\%). For comparison, we also include the disk read volume of our {\proposed}, using a fixed 32 MB buffer (less than 1\% of the graph), shown as horizontal lines in \figref{fig:overhead1}.

\header
\textbf{Read Inflation.} 
Now consider the first source of I/O inefficiency in synchronous execution: \textit{read inflation}. Although BFS accesses each edge exactly once, SSDs operate at block granularity—so accessing a single edge requires loading its entire block. Many of these edges may be unused, resulting in unnecessary reads. Worse, the limited cache size may evict a block before all its edges are accessed, leading to redundant reloads, which we call it as \textit{read inflation}. For example, if block~$B$ contains edges for vertices $v_1$ and $v_2$, and $v_1$ is activated in Iteration~1 while $v_2$ is activated in Iteration~3, block~$B$ may be evicted and must be reloaded—resulting in duplicated I/O.

A larger cache pool can alleviate the issue of read inflation to some extent by holding more blocks, increasing the chance of cache hits for repeated accesses. In the extreme case where the entire graph fits in memory, BFS incurs no read inflation. However, as noted in the introduction, real-world graphs are rapidly growing and often exceed a single machine’s memory capacity.

As shown in \figref{fig:overhead1}(a)--(b), all cache replacement strategies reduce disk read volume as the buffer pool increases from 4 MB to 20\% of the graph size. For instance, the OPT strategy reduces reads by 29\% on Twitter and 25\% on Friendster. OPT consistently outperforms the others due to its idealized knowledge of future accesses. However, even with a buffer size equal to 20\% of the graph size, OPT still performs worse than our asynchronous {\proposed} on Friendster using only a 32 MB buffer, highlighting the limitations of synchronous execution under read inflation.

\header{\bf Work Inflation.} The WCC experiments reveal the second source of I/O inefficiency in synchronous graph processing systems: work inflation. In parallel WCC computation, the common approach is to use the Label Propagation (LP) algorithm, which initializes each vertex with a unique label (typically its vertex ID) and iteratively updates each vertex’s label to the minimum label among its neighbors until global convergence is achieved. In a sequential LP algorithm, the propagation can be focused by starting from the vertex with the minimum label and selectively propagating to its neighbors. In contrast, the synchronous setting propagates updates across all vertices in each iteration to accelerate convergence, a strategy that often leads to unnecessary computations compared to the more targeted sequential approach, thereby resulting in work inflation. This work inflation further causes unnecessary I/Os. For example, on both the Twitter and Friendster graphs, over $99\%$ of I/O costs stem from the first 2–3 iterations, during which nearly all vertices remain active. \reviseROne{This forces almost the full graph to be reloaded into memory each iteration, limiting the effectiveness of cache policies.} As shown in Figs. \ref{fig:overhead1} (c)-(d), disk read reduction scales near-linearly with cache pool size. \reviseROne{Even with idealized policies (\textit{OPT}, \textit{SUB}) or practical ones (\textit{LRU}), the performance gap between strategies is minimal (under 0.7\% on both graphs), and achieve only $10\%$ I/O reduction with 20\% buffer pool size.} Consequently, relying solely on the buffer pool is insufficient to further mitigate the I/O costs arising from work inflation in synchronous GPSs. \reviseROne{Actually, within a connected component, only updates from vertices with the smallest label to their higher-labeled neighbors are effective. All other label updates will finally be overwritten and thus redundant. By prioritizing blocks containing active vertices with the minimum label, {\proposed} enables the algorithm to execute in a better order, thereby reducing redundant edge accesses and computations, leading to reduced disk I/O.
}

\begin{figure}[!t]
\begin{small}
    \begin{tabular}{c@{\hspace{10mm}}c@{\hspace{10mm}}c}
    \multicolumn{3}{c}{\hspace{-3mm}\includegraphics[width=0.5\linewidth]{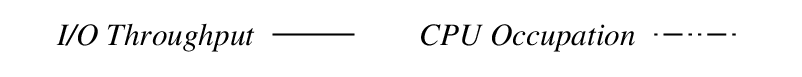}}\\
    \includegraphics[height=22mm]{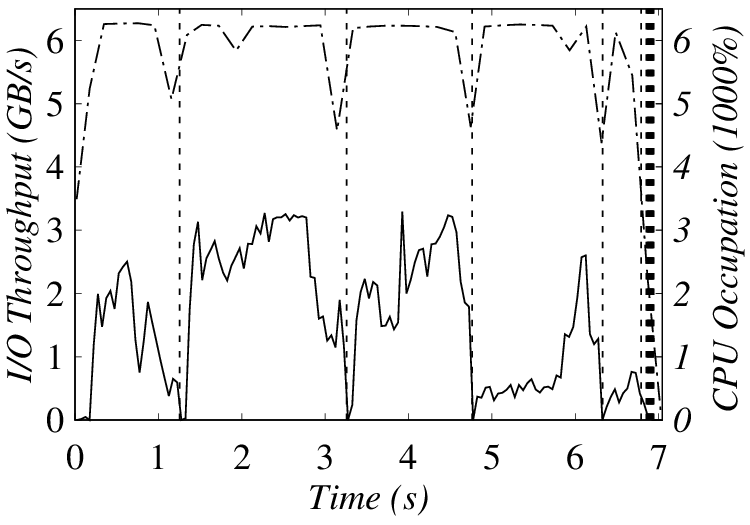} &
    \includegraphics[height=22mm]{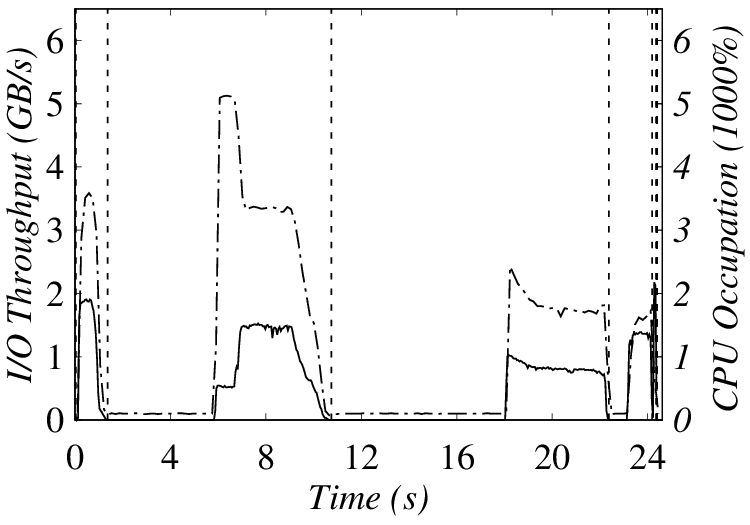} &
    \includegraphics[height=22mm]{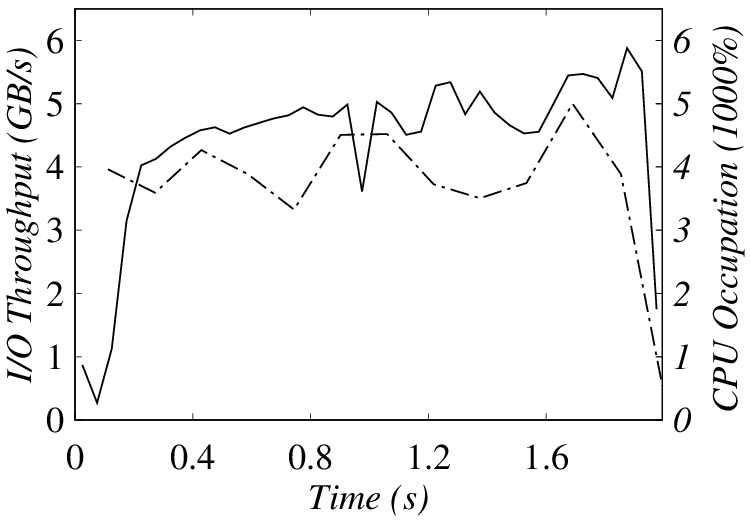} \\
    (a) Blaze & (b) CAVE & (c) ACGraph (ours) \\
    \end{tabular}
    \vspace{-1mm}
    \caption{\reviseROne{Disk throughput \& CPU occupation over time on BFS queries; Vertical dashed lines mark the starting point of each iteration.}}
    \label{fig:overhead2}
    \vspace{-4mm}
\end{small}
\end{figure}

\subsection{Synchronization Stalls} \label{sec:overhead2}
\reviseROne{\figref{fig:overhead2} presents the I/O throughput and CPU occupation over time across graph processing frameworks \cite{Blaze, CAVE} during BFS execution on the Twitter graph using 64 threads.} We have omitted the system initialization and graph index loading phases, with time zero representing the actual start of the algorithm. The IO throughput curves are obtained through disk sampling using the \texttt{blk\_trace} \cite{blktrace} tool on Linux, with a 50 ms sampling window, where throughput is calculated based on the number of complete actions of read operations within each window. \reviseROne{Both Blaze and CAVE exhibit periodic throughput and CPU drops around each iteration, which we refer to as \textbf{synchronization stalls}.} Notably, CAVE even experiences prolonged zero I/O at the beginning of each iteration, as it employs a single thread for result collection, task generation, and distribution. \reviseROne{A more critical challenge arises from ``log-tail'' iterations: In large-diameter graphs (e.g. Uk-Union), algorithms like BFS may run for hundreds of iterations, with over $80\%$ of them activating fewer than 1,000 vertices. This lead to low average throughput despite high peak performance. Hence, even though both systems utilize work-stealing to reduce load imbalance, Blaze's average throughput is only 2.4~GB/s (compared to a 4.1~GB/s peak), and CAVE's is merely 0.27~GB/s (compared to a 2.3~GB/s peak).}

The inefficiencies inherent in synchronous out-of-core GPSs as mentioned above motivate us to introduce ACGraph, an asynchronous graph processing system. In the following section, we detail our design of the ACGraph framework.

\section{The ACGraph Framework} \label{sec:framework}
In this section, we present the details of the ACGraph framework. In ACGraph, edges are partitioned into 4KB blocks and stored on SSD, while vertex data (e.g., vertex states, degrees, and edge offsets) are maintained in memory. ACGraph ensures that the edges of a vertex span only one block if they can fit into 4KB. If the edges of a vertex are partitioned into a block, we logically assign the vertex to that block, hereafter referred to as the vertex's \textit{assigned block} or \textit{associated block}. Graph partitioning strategies and corresponding storage optimizations are discussed in \secref{sec:storage}.

\subsection{Block-Centric Async.\ Execution Model}
The challenges outlined in \secref{sec:motivation} present significant resolution difficulties under synchronous semantics or stem from inherent limitations of the synchronous paradigm. This motivates our exploration of an asynchronous execution model. Without the synchronization constraints, when a block is loaded, processing can continue as long as it contains active vertices, thereby improving resource utilization and reducing read inflation. While not every vertex within a scheduled block may contribute optimal progress, this overhead is negligible against the substantial reduction in I/O operations. In addition, block execution order can be priority-scheduled to enhance work efficiency for certain algorithms (e.g., WCC). Furthermore, the asynchronous model only requires a single global synchronization at the algorithm's conclusion, eliminating per-iteration synchronization overhead and associated load imbalance issues.

\begin{figure}
    \centering
    \includegraphics[height=18mm]{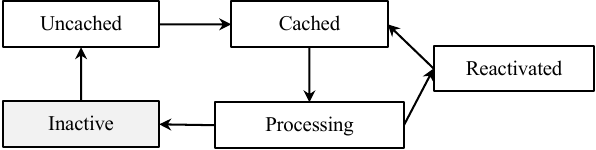}
    \vspace{-1mm}
    \caption{Block state machine.}
    \label{fig:bstate}
   \vspace{-2mm}
\end{figure}

Building on these principles, we propose a novel priority-enabled asynchronous execution model designed for modern SSDs: \textbf{the block-centric execution model}. The block-centric execution model, as what the name suggests, explicitly treats blocks as first-class entities rather than mere disk-stored data, as seen in traditional out-of-core systems, while vertices are associated with their assigned blocks, instead of being treated as standalone entities.
To clarify terminology, a \textit{block} in ACGraph refers to a logical scheduling unit containing a disjoint partition of vertices. A block corresponds to an SSD region termed a \textit{disk block}, which becomes \textit{block data} when loaded into memory.
Each block maintains metadata including status, priority, active vertices in it, and a pointer referencing its block data. 
Scheduling decisions operate at the block level, where block priorities are derived by aggregating the priorities of their frontiers (e.g., using the maximum value for high-priority-first or minimum value for low-priority-first). Individual vertex priorities, in contrast, are application-defined—for instance, using vertex distance as the priority metric in BFS.
Every active vertex is maintained by the local frontier set of its assigned block, instead of a global one. When the priority of a vertex $v$ gets updated, the priority of its associated block is also updated.

Each block in this execution model works as a state machine as shown in \figref{fig:bstate}. Except \textit{Inactive}, the  remaining four states indicate that a block contains active vertices at different execution phases:
\begin{itemize}[leftmargin=*]
    \item \textit{Uncached}: Block data has not been loaded into memory.
    \item \textit{Cached}: Block data is loaded in memory but not yet processed.
    \item \textit{Processing}: The block is being processed by an executor.
    \item \textit{Reactivated}: Some vertices in the block are newly activated during processing.
\end{itemize}
The state of each block dynamically evolves. All blocks initialize in an inactive state. When a vertex is activated, the associated block transits to an uncached state. When memory is available, the system prioritizes and loads the uncached block with the highest priority into memory, changing its state to cached for execution readiness. A cached block enters the processing state when being executed. If new vertex activations occur during processing, the block is marked as reactivated. After the processing of the block, reactivated blocks immediately return to the cached state for prioritized re-execution, while blocks without reactivation revert to the inactive state. Under this design, an in-memory subgraph is dynamically maintained by active blocks. Vertices within the in-memory subgraph are processed preferentially once activated.

\subsection{The Architecture of ACGraph}

\begin{figure}[t]
    \centering
    \hspace{-2mm}
    \includegraphics[height=58mm]{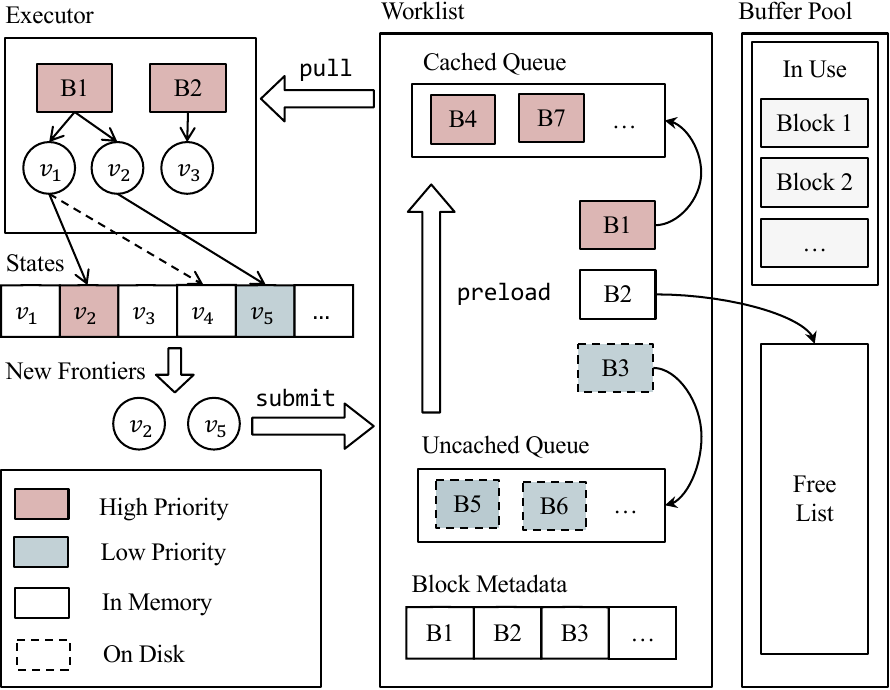}
    \vspace{-1mm}
    \caption{The architecture of ACGraph (Bi indicates block i).}
    \label{fig:arch}
    \vspace{-3mm}
\end{figure}

\figref{fig:arch} shows the architecture of ACGraph, comprising three core components: \textit{executors}, \textit{worklist}, and \textit{buffer pool}. Each executor runs on a dedicated thread and processes tasks as described in \algref{algo:executor}. Executors continuously interact with the worklist—the system’s scheduler—to pull and submit tasks. The worklist dictates execution order and interleaves task scheduling with I/O using \texttt{io\_uring} to avoid I/Os blocking executors. It also maintains block metadata and manages state transitions. The buffer pool maintains and recycles a fixed set of memory buffers to improve access efficiency.

\header
\textbf{Executor.} 
The visualization in \figref{fig:arch} employs color (red for high-priority, blue for low-priority blocks) and border style (solid for memory-resident, dashed for disk-resident) to represent block states. \algref{algo:executor} outlines the executor's workflow. The executor continuously pulls a batch of tasks from the worklist until it is exhausted (Line 1), where a task corresponds to a block and includes a pointer to preloaded block data and a set of active vertices (Line 4). For example, the executor in \figref{fig:arch} acquires two block tasks, where block 1 contains active vertices $v_1$ and $v_2$, while block 2 has only one frontier $v_3$. 
The executor iteratively processes the active vertices of each task (Lines 2-5). The adjacency lists of the vertices in a block task are stored in separate locations within the block data, which can be located via the vertex information maintained in memory (Line 6). The executor processes these vertices by first calling a user-defined $\mathtt{apply}$ function to generate messages for their neighbors (Line 7). Then it updates their neighbors' states using a user-defined $\mathtt{propagation}$ function (Line 13). Newly activated vertices ($v_2$ and $v_5$) are temporarily stored in a local buffer (Lines 14-15), which is submitted to the worklist when necessary (Line 9). These vertices are grouped by their assigned blocks to update corresponding metadata in the worklist ($v_2$ is grouped to block 1 while $v_5$ to block 3). Upon completing a block’s processing, the executor invokes the $\mathtt{finish}$ method to notify the worklist and update the block’s state (Line 10).

\header 
\textbf{Worklist.} The worklist serves as the core data structure of ACGraph and is pivotal for enabling sustained computation and I/O triggering. The worklist maximizes block data reuse through two components: a queue for scheduling loaded blocks and a priority queue for scheduling unloaded blocks. This dual-queue architecture ensures memory-resident blocks always precede disk-resident ones in execution order. This prioritization stems from the $\mathtt{pull}$ operation exclusively retrieving batched blocks from the cached queue. During task submission, the worklist dynamically adapts to block states: uncached blocks such as block~3 are routed to the uncached queue, while cached blocks enter the cached queue. For blocks in processing or reactivated states (exemplified by block~1), the system updates their frontier metadata and priority values, transitioning their state to reactivated without queue insertion. The $\mathtt{finish}$ method initiates state-specific transitions: reactivated blocks like block~1 re-enter the cached queue, while inactive blocks release allocated buffer pool resources, as demonstrated by block~2. This state-aware design guarantees execution priority through three operational principles: cached queue dominance in task retrieval, state-driven transitions for active blocks, and automated memory reclamation via buffer pool integration.

\header 
\textbf{Buffer Pool.} The buffer pool pre-allocates memory space for storing block data and employs a concurrent queue (the free list shown in \figref{fig:arch}) to manage available slots. It supports batch allocation and release operations for block data through a single call. Notably, the buffer pool itself does not provide data query functions; whether a specific block data resides in the pool is ultimately determined by external metadata queries tied to each block.

\begin{algorithm}[t]
\caption{Workflow of executors}
\label{algo:executor}
\BlankLine
\KwIn{Input graph $G=(V,E)$, worklist $W$, user-defined function \texttt{apply} and \texttt{process}}
\SetKwProg{Fn}{Function}{}{}
\SetKwFunction{propagation}{propagation}
\SetKwFunction{Pull}{pull}
\SetKwFunction{Submit}{submit}
\SetKwFunction{Finish}{finish}
\SetKwFunction{Apply}{apply}
\SetKwFunction{Process}{process}
\SetKwFunction{Append}{append}
\SetKwFunction{GetAdj}{getNeighbors}
 \SetAlgoNoEnd
\BlankLine

\While{ $tasks \gets W.\texttt{pull}()$} {
    \For{$task \in tasks$} {
        init an empty list $buffer$\;
        $tid,frontiers, data \gets task$ \;
        \For {$u \gets frontiers$} {
            $neighbors \gets G.$\GetAdj{u, data} \;
            $msg \gets $ \Apply{u} \;
            \Process{msg, neighbors, buffer};
        }
        $W.$\Submit{buffer}\;
        $W.\texttt{finish}(tid)$ \;
    }
}

\Fn{\Process{$msg, neighbors, buffer$}} {
    \For{$v \gets neighbors$} {
        $priority \gets $\propagation{msg, v}\;
        \If {$priority > 0$} {
        $buffer.$\Append{priority, v};
        }
    }
}
\end{algorithm} 

\reviseROne{
\subsection{Synchronous Execution} \label{sec:sync}
Asynchronous execution suits algorithms that converge rapidly under arbitrary ordering, such as $k$-core. Another class of problems requires specific execution orders for optimal time complexity yet converges correctly under relaxed orders (e.g., WCC). {\proposed} supports such cases, delivering correct results while leveraging priority scheduling to enhance practical performance.

However, some algorithms fundamentally require global synchronization for correctness. Consider maximal independent set (MIS) problem: an efficient parallel algorithm (i.e. Blelloch’s Algorithm 2 \cite{MIS}) randomly assigns each vertex a fixed unique label, initially marks all vertices as live and create an empty set $s$. Each iteration adds the live vertices with no lower-labeled live neighbors to set $s$, then marks them and their neighbors as dead. This repeats until all vertices become dead, and set $s$ forms a valid MIS.

Fortunately, synchronous execution is a special case of asynchronous execution. {\proposed} supports synchronous mode with minimal modification: new frontiers are inserted into a fresh worklist instead of the current one. This creates global barriers, ensuring that newly activated vertices are only processed after all current frontiers have completed. With this flexibility, users can select the execution mode best suited to their algorithm's requirements.

Note that {\proposed} does not automatically select execution modes. Instead, it exposes distinct APIs for users to match specific requirements (see \secref{sec:api}). While adaptive execution mode selection based on runtime conditions represents an active research area \cite{XieCGZC15, adaptiveasync}, this capability remains beyond our current scope. We therefore designate automated execution mode switching on out-of-core graph processing as promising future work.
}

\reviseROne{
\subsection{Consistency and Correctness} \label{sec:consistency}
{\proposed} permits concurrent vertex/edge access by multiple threads. To ensure consistency, users must implement appropriate concurrent control (e.g., atomic variables or locks) in user-defined functions (see \secref{sec:example}), as in prior frameworks \cite{CAVE, ligra, galois}. This offers two benefits: first, it avoids system-level locking overhead or additional scheduling constraints that could hinder parallelism (e.g., preventing simultaneous execution of vertices with shared neighbors); second, it enables users to balance performance and consistency via selecting concurrency control primitives. For example, in scenarios permitting precision loss, one may trade consistency and correctness for higher performance by employing lightweight atomic operations (e.g. replacing \texttt{compare\_and\_swap} with a simple \texttt{store}) and relaxed memory ordering.}

\reviseRTwo{With proper user-level implementations, such as atomic and data-race-free \texttt{apply} and \texttt{propagation} functions, {\proposed} achieves \textit{sequential consistency}, meaning its results correspond to some valid sequential order. Thus, if the algorithm is correct under all sequential executions, it remains correct on {\proposed}. This matches the consistency level of existing in-memory asynchronous GPSs, such as GraphLab \cite{graphlab}. In asynchronous mode, {\proposed} ensures that a vertex is processed only after it is activated, and the number of invocations never exceeds its activation count. These properties help verify algorithm correctness (see \secref{sec:example} for example).
}

\begin{figure}[t]
    \centering
    \hspace{-2mm}
    \includegraphics[width=0.55\linewidth]{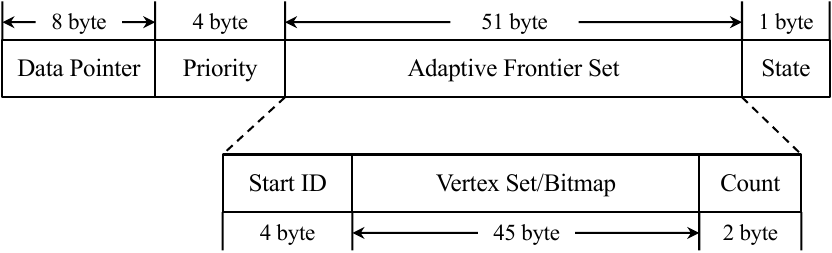}
    \vspace{-1mm}
    \caption{Block metadata.}
    \label{fig:metadata}
   \vspace{-2mm}
\end{figure}

\subsection{Block Management}
\header \textbf{Block Metadata.}
The block metadata structure (\figref{fig:metadata}) uses 64-byte alignment to reduce false sharing in concurrent access. It consists of four parts: an 8-byte atomic data pointer, a 4-byte priority field encoded as a signed integer, a 51-byte adaptive frontier set (AFS), and a 1-byte state flag. Mutual exclusion is implemented using the lowest bit of the data pointer as a spinlock, avoiding dedicated mutexes. The 51-byte AFS is organized into three segments: a 4-byte Start ID $v_{start}$ which marks the smallest vertex ID in the block, a 2-byte counter to track the number of active vertices, and a 45-byte storage space. The AFS implements dual storage strategies for memory and computation efficiency. In sparse mode, active vertices are directly stored in an array with a maximum capacity of $\lfloor 45/4\rfloor = 11$ vertices (4-byte per vertex). The dense mode uses the 45 bytes (360 bits) as bitmap encoding, to represent vertices' activity within the range $[v_{start}, v_{start}+360)$. Mode transitions occur dynamically based on vertex count, with $v_{start}$ enabling consistent offset calculations.

A 4KB disk block can hold up to 1,024 edges. Extreme cases with all degree-1 vertices would require 1,024 bits (128 bytes), exceeding metadata limits. The hybrid storage solution (\secref{sec:storage}) resolves this by keeping vertices with degree $\leq 2$ in memory, and retaining only those with degree $\geq 3$ within blocks, without incurring additional memory overhead. This reduces the theoretical maximum vertex capacity to  $\lfloor 1024/3\rfloor < 342$, while the AFS supports $360$ vertices in the dense model (using bitmaps), which is sufficient.

\header
\textbf{Preload.} Whenever an executor attempts to pull tasks from the worklist (Algorithm \ref{algo:executor} Line 1),  it first triggers a \textit{preload} process for other blocks. This mechanism retrieves a batch of highest-priority blocks from the uncached\_queue, allocates buffers from the buffer pool if the buffer pool is not full, and generates asynchronous I/O tasks to read disk blocks. The preload process also manages synchronization of completed I/O tasks, updates the state of these blocks to cached, and delegates them to the cached\_queue for scheduling. Disk I/O is executed asynchronously via io\_uring, enabling threads to collect completed I/O operations (freeing resources) in  
a non-blocking manner, submit new I/O tasks, and return immediately without waiting. ACGraph avoids segregating I/O and computation threads, as it employs adaptive thread distribution. For example, under a “fast computation, slow I/O” scenario, the executor increases its task-pulling frequency, thereby issuing more asynchronous I/O requests to saturate disk bandwidth and maximize throughput.

\reviseROne{
\header
\textbf{Early-stop.}
A concern regarding the reuse of reactivated blocks is potential computational inefficiency, as this approach may disrupt the priority-based block execution order. For example, in BFS algorithms, blocks retained in memory might form a suboptimal long path, rendering many subsequent updates redundant. To mitigate this, {\proposed} employs an early-stop strategy: any block reused consecutively beyond a user-set threshold is forcibly evicted from memory and returned to the uncached queue for scheduling. Notably, our experiments do not reveal significant effect of this issue. Consequently, ACGraph disables this strategy by default.
}

\subsection{APIs} \label{sec:api}
ACGraph abstracts implementation details via two high-level APIs: $\mathtt{foreachVertex}$ and $\mathtt{asyncRun}$, \reviseROne{which follow the design philosophy of Ligra \cite{ligra}: The first supports parallel traversal of vertices (analogous to VERTEXMAP), while the second iterates over the edges of the frontier (similar to EDGEMAP). This conceptual alignment enhances both usability and simplicity. We adapt their semantics to better support asynchronous execution. Specifically, the API $\mathtt{foreachVertex}$ iterates over all vertices, inserting activated vertices back into the given worklist rather than creating a new one. Meanwhile, the API $\mathtt{asyncRun}$ repeatedly processes active vertices in the worklist and reinserts newly activated vertices, continuing this cycle until no active vertices remain.}


The definition of $\mathtt{foreachVertex}$ is listed in Eqn.\ \eqref{eq:foreach}, with three parameters: a graph $G$, an optional worklist $W$, and a vertex-to-priority function $f$. When invoked, $\mathtt{foreachVertex}$ applies $f$ to all vertices of $G$ in parallel. If the return value of $f$ for a vertex is greater than $0$, the vertex is submitted to the worklist $W$ (if provided), indicating it is activated. \texttt{foreachVertex} is ideal for initializing vertex states and generating initial frontiers efficiently.
\begin{equation}
\label{eq:foreach}
    \begin{aligned}
        \mathtt{foreachVertex}(&G:  \mathtt{graph}, W: \mathtt{worklist}, \\
        &f: ( \mathtt{vertex})\rightarrow  \mathtt{priority}) \rightarrow  \mathtt{void}
    \end{aligned}
\end{equation}

The $\mathtt{asyncRun}$ method, as abstracted in Eqn. \eqref{eq:asyncrun}, takes four parameters: a graph $G$, a worklist ${W}$ of initial active vertices, and two functions $\mathtt{apply}$ and $\mathtt{propagation}$. During execution, $\mathtt{asyncRun}$ processes each active vertex in ${W}$ by first invoking the $\mathtt{apply}$ function to generate messages destined for its neighbors. These messages are then propagated to neighboring vertices via the $\mathtt{propagation}$ function. If the return value of $\mathtt{propagation}$ for a neighbor exceeds zero, the neighbor is marked as activated and inserted into the worklist ${W}$, ensuring iterative updates until  convergence.
\begin{equation}
\label{eq:asyncrun}
    \begin{aligned}
         \mathtt{asyncRun}(&G:  \mathtt{graph},\, W:  \mathtt{worklist},\, \mathtt{apply}: ( \mathtt{vertex})\rightarrow  msg\_t,\, \\
        & \mathtt{propagation}: ( msg\_t,  \mathtt{vertex})\rightarrow  \mathtt{priority} )\rightarrow  \mathtt{void}
    \end{aligned}
\end{equation}

\reviseROne{Additionally, {\proposed} provides a synchronous $\mathtt{syncRun}$ API. Its interface resembles $\mathtt{asyncRun}$ but aggregates newly activated vertices into a distinct worklist , which is returned upon completion.}

\subsection{Example}\label{sec:example}

\begin{algorithm}[t]
\caption{Breadth-First Search}
\label{algo:bfs}
\BlankLine

\SetKwProg{Fn}{function}{}{}
\SetKwFunction{propagation}{propagation}
\SetKwFunction{Pull}{pull}
\SetKwFunction{Submit}{submit}
\SetKwFunction{Finish}{finish}
\SetKwFunction{Apply}{apply}
\SetKwFunction{Process}{process}
\SetKwFunction{Append}{append}
\SetKwFunction{GetAdj}{getNeighbors}
\SetKwFunction{CAS}{CAS}
\SetKwFunction{AsyncRun}{asyncRun}
\SetAlgoNoEnd

\KwIn{Input graph $G=(V,E)$, user-defined function \texttt{apply} and \texttt{process}, source vertex $s$}
\BlankLine
$dis \gets \{\infty, \dots, \infty\}$\;
$dis[s] \gets 0$\;
Initialize a worklist $W$\;
$W.\Submit{s}$\; \
$\AsyncRun{G, W, \Apply, \propagation}$\;

\BlankLine

\Fn{\Apply{$u$}} {
    \Return {$dis[u]$.\texttt{load()}}\;
}

\BlankLine
\Fn{\propagation{$msg, v$}} {
    $d \gets dis[v].\texttt{load()}$\;
    \While{d > msg+1} {
        \If {$dis[v].$\CAS{d, msg+1}} {
            \Return{msg+1}\;
        }
        $d \gets dis[v].\texttt{load()}$\;
    }
    \Return{0}\;
}
\end{algorithm} 

We demonstrate the versatility of these APIs through two canonical examples: BFS (Breadth-First Search) and $k$-core.
In our example, atomic variables are employed to enforce thread safety. Two atomic operations are utilized: $\mathtt{load}()$ reads the current value of a variable. The $\mathtt{CAS(old, new)}$ operation compares the variable’s current value with $\mathtt{old}$; if equal, updates it to $\mathtt{new}$ and returns true; otherwise, returns false without modification. The $\mathtt{fetchSub}(x)$ operation atomically subtracts ${x}$ from the variable’s value and returns its original value. 
\revise{All operations are performed atomically, utilizing \texttt{std::memory\_order\_acquire} for \texttt{load} and \texttt{std::memory\_order\_acq\_rel} for \texttt{CAS} and \texttt{fetchSub} to establish acquire-release semantics. This guarantees that all memory writes before a given release operation become visible to subsequent acquire operations on the same variable.}


\header
\textbf{BFS.} \algref{algo:bfs} implements BFS using the offered APIs. Lines 1-2 initialize a distance array $dis$ of size $|V|$ with all values set to $\infty$ representing unvisited vertices, except the source vertex $s$'s distance to $0$. Lines 3-4 create an empty worklist $W$ and activate the source vertex. The ACGraph framework then autonomously executes message passing via its $\mathtt{asyncRun}$ method until convergence (when the worklist processed by the executor becomes empty). The $\mathtt{apply}$ function (Lines 6-7) directly returns the vertex's current distance as messages. The $\mathtt{propagation}$ function (Lines 8-14) compares each neighbor $v$'s distance with the candidate value $msg+1$.  If $msg+1$ is smaller, it attempts to atomically update $dis[v]$ using a CAS operation.  Successful updates activate neighbor $v$ with priority $msg+1$ via insertion into $W$, propagating the BFS frontier iteratively. 

\revise {
The correctness of \algref{algo:bfs} can be proved by mathematical induction. Assume that all vertices with $dis[\cdot] = d$ (for any $d \geq 0$) have correctly converged to their distances. Now consider a vertex $v$ with distance $d+1$. By definition, $v$ must be a neighbor of some vertex $u'$ with distance $d$. The \texttt{propagation} function ensures that when $dis[u']$ is updated to $d$, $u'$ is activated. {\proposed} assures that once activated, $u'$ will be processed, triggering \texttt{propagation} and allowing $dis[v]$ to be updated based on $dis[u']$. As a result, $dis[v]$ converges to $d+1$. Since source $s$ is correctly initialized with distance 0 and activated, this inductive process ensures that all vertices reachable from $s$ will finally converge to their correct distances.
}

\begin{algorithm}[t]
\caption{$k$-core}
\label{algo:kcore}
\BlankLine

\SetKwProg{Fn}{function}{}{}
\SetKwFunction{propagation}{propagation}
\SetKwFunction{Pull}{pull}
\SetKwFunction{Submit}{submit}
\SetKwFunction{Finish}{finish}
\SetKwFunction{Apply}{apply}
\SetKwFunction{Init}{init}
\SetKwFunction{GetDegree}{getDegree}
\SetKwFunction{Process}{process}
\SetKwFunction{Append}{append}
\SetKwFunction{GetAdj}{getNeighbors}
\SetKwFunction{CAS}{CAS}
\SetKwFunction{AsyncRun}{asyncRun}
\SetKwFunction{ForeachVertex}{foreachVertex}
\SetKwFunction{FetchSub}{fetchSub}
\SetAlgoNoEnd

\KwIn{Input graph $G=(V,E)$, user-defined function \texttt{apply} and \texttt{process}, threshold $k$}
\BlankLine
$deg \gets \{0, \dots, 0\}$ \;
Initialize a worklist $W$\;
$\ForeachVertex{G, W, \Init}$\;
$\AsyncRun{G, W, \Apply, \propagation}$\;

\BlankLine

\Fn{\Init{$u$}} {
    $deg[u] \gets G.\GetDegree{u}$\;
    \If {$deg[u] < k$} {
    \Return{1}\;
    }
    \Return{0}\;
}

\BlankLine

\Fn{\Apply{$u$}} {
    \Return {0}\;
}

\BlankLine
\Fn{\propagation{$msg, v$}} {
    $d \gets deg[v].$\FetchSub{1} \;
    \If {d = k} {
    \Return{1}\;
    }
    \Return{0}\;
}
\end{algorithm} 

\header
\textbf{$\boldsymbol{k}$-core.} The $k$-core algorithm iteratively prunes vertices with degree less than $k$ and their incident edges until no such vertices remain. We model vertex states using their degrees, constructing a global array $deg$ to store these values. Unlike BFS, Line 3 leverages the $\mathtt{foreachVertex}$ method to parallelize initialization: the $\mathtt{init}$ function (Lines 5-9) sets each vertex’s state to its initial degree, while vertices with $deg[v]<k$ are activated and added to the worklist (Lines 7-8). During the propagation phase, the algorithm decrements the target vertex’s degree by $1$ to simulate edge removal (Line 13). Following this update, if the vertex’s degree equals $k$ exactly before the update, it is activated to propagate further pruning (Lines 14-15). The equality constraint ensures that each vertex is activated at most once, guaranteeing algorithmic correctness. This process continues until convergence, i.e., when the worklist processed by the executor becomes empty.

\revise {
The correctness of \algref{algo:kcore} can be established by mathematical induction also. Consider a valid vertex removal sequence $v_1, v_2, \dots, v_t$ produced by the serial $k$-core algorithm. Suppose vertex $v_1$ through $v_i$ have already been correctly activated in \algref{algo:kcore}. For $v_{i+1}$, if its initial degree is less than $k$, it is activated in \texttt{init}. Otherwise, there exists a in-neighbor $u \in \{v_1, \dots, v_i\}$ of $v_{i+1}$ whose activation triggers a \texttt{propagation} operation that reduces $\text{deg}[v_{i+1}]$ to $k-1$ (guaranteed by the atomicity of \texttt{fetchSub}), leading to the correct activation of $v_{i+1}$. Since $v_1$ always has initial degree less than $k$, it is correctly activated in \texttt{init}. Besides, {\proposed} ensures that each activated vertex executes exactly once in this context, preventing erroneous activation of $k$-core vertices. Therefore, a vertex is activated in \algref{algo:kcore} if and only if it lies outside the $k$-core.

}
\section{Hybrid Storage Architecture} \label{sec:storage}

\begin{figure}[t]
    \centering
    \includegraphics[width=0.6\linewidth]{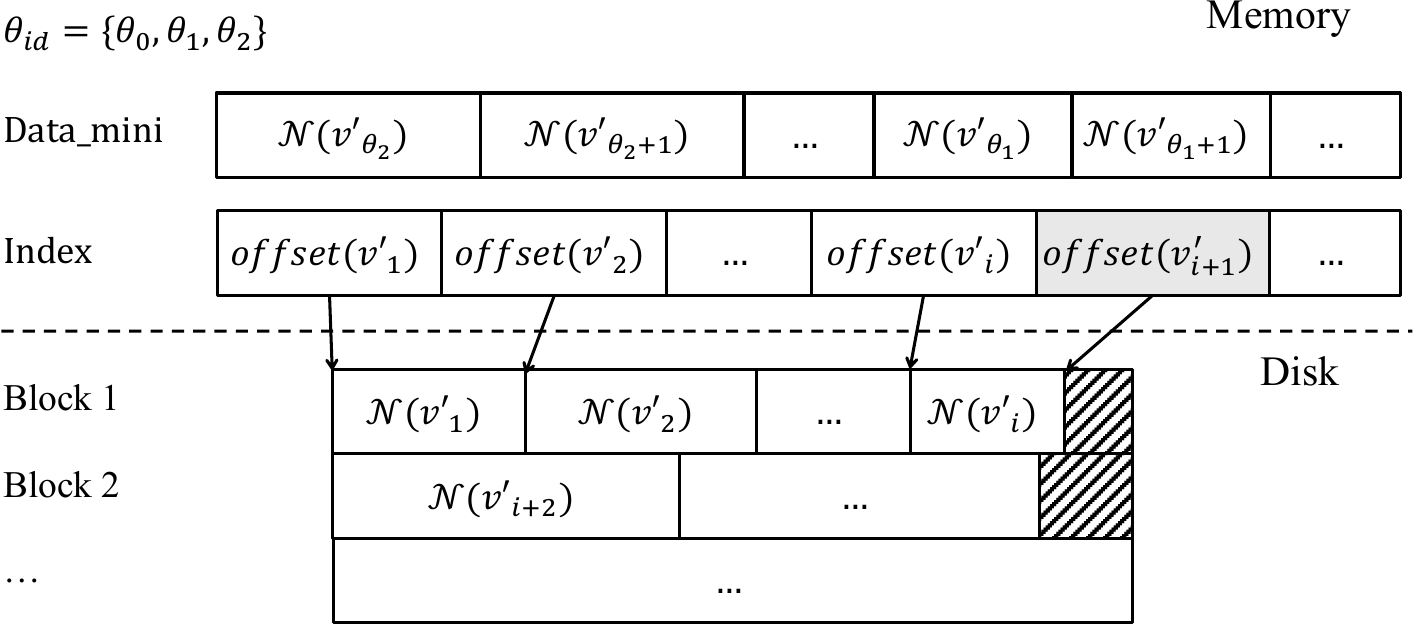}
    \vspace{-1mm}
    \caption{The hybrid graph storage in ACGraph, with $\degt=2$, where $v_i'$ is the $i$-th vertex after vertex reordering and the gray cell represents virtual vertices. }
    \label{fig:storage}
   \vspace{-2mm}
\end{figure}

In in-memory GPSs, \textit{Compressed Sparse Row (CSR)} is widely used for graph storage. The CSR structure has two core arrays: the \textit{index} array storing offsets to adjacency lists in the \textit{data} array, and the \textit{data} array containing edges, which stores only destination vertices. 
As the adjacency list is closely arranged in \textit{data}, vertex degrees can derive directly from offset differences via $\mathtt{deg}(v) = \mathtt{offset}(v+1) - \mathtt{offset}(v)$, avoiding extra degree storage. 

\reviseROne{
Existing semi-external GPSs present several techniques to optimize storage. Blaze \cite{Blaze} groups sixteen 4-byte degrees per cache line and stores only cache-line locations in the offsets region. While this saves memory by removing offset fields for vertices, it introduces additional prefix-sum computation overhead for indirect vertex offset access. CAVE \cite{CAVE} uses 4-byte offsets to reduce memory, but this limits its scalability for large graphs. ChunkGraph \cite{chunkgraph} embeds mini edge lists into offsets, but still requires a separate degree array to identify vertex types. In contrast, we propose a hybrid storage architecture that eliminates the degree field and allows mini adjacency lists (for small-degree vertices) to reside directly in memory without extra metadata, while supporting block-centric execution (see \figref{fig:storage}). The in-memory \textit{index} array and vertex states (not shown in \figref{fig:storage}) leverage fast random RAM access, which now can hold billions of vertices on standard DRAM. To support block-centric execution, {\proposed} enforces that adjacency lists smaller than 4KB fit within a single block, while larger ones span consecutive blocks.}

To achieve this, {\proposed} employs a locality-preserving last-fit partition policy strategy (\secref{sec:partition}), while preserving inherent locality \cite{galois}. Yet, these constraints disrupts CSR's degree-offset property, i.e., 
\(
\mathtt{deg}(v) = \mathtt{offset}(v+1) - \mathtt{offset}(v).
\)
This disruption occurs for two reasons:  {\em (i)} Non-monotonic offsets, as now it is possible for \(\mathtt{offset}(v) > \mathtt{offset}(u)\) even when \(v < u\); {\em (ii)} Internal block fragmentation, as not all blocks are fully utilized. Hence, each vertex must store a 12-byte index (8-byte for the offset and 4-byte for the degree), which may expand to 16-byte due to memory alignment when stored in a \texttt{struct}, inflating memory costs. We eliminate the overhead with two simple yet effective techniques: \textit{vertex reordering} and \textit{virtual vertex insertion}. These allow us to store mini adjacency lists directly in memory without additional overhead. Together, these optimizations (see \secref{sec:space-opt}) form the cornerstone of ACGraph’s hybrid storage.

\subsection{Graph Partition Strategy} \label{sec:partition}

Partitioning graphs into 4KB blocks under the specified constraints introduces a trade-off between preserving graph locality and minimizing intra-block fragmentation. For example, a best-fit approach that minimizes fragmentation by distributing vertices to blocks with an exact fit may scatter low-degree vertices across multiple blocks. This scattering undermines inherent locality and is particularly disadvantageous for algorithms, such as $k$-core, that primarily access small vertices. On the other hand, partitioning vertices without reordering may introduce large fragmentation across many blocks, resulting in severe read inflation issues.

To balance these factors, ACGraph employs a heuristic locality-preserving last-fit partitioning strategy. Specifically, only blocks within a sliding window are considered as candidate placements for the current vertex. When no block in the window can accommodate the vertex's edge list, new blocks are created and the window shifts accordingly. Adjusting the window size allows for a tunable trade-off between locality preservation and fragmentation control.

This strategy can be implemented via a segment tree or direct sliding window traversal (efficient for small window sizes).  Both approaches iterate over vertices while tracking allocated blocks. In the segment tree approach, each tree node stores the maximum remaining space across a range of blocks. For each vertex, the system queries the rightmost block that can accommodate the vertex's edge list. If no suitable block exists within the window, a new block is created. The vertex is then inserted into the selected/created block, with subsequent updates to both the sliding window and segment tree. For small sliding windows, a direct traversal identifies the rightmost available block that meets the space requirements. We use a default window size of $8$ and the direct window traversal method in our system to prioritize spatial locality.

We further enhance efficiency through multi-threading, assigning each thread a subset of vertices to process. Although this parallelism may result in a slight increase in intra-block fragmentation, it significantly reduces preprocessing time. Notably, because only vertex degrees are required for this strategy, it remains practical in memory-constrained environments.

\subsection{Space Optimization}\label{sec:space-opt}

We implement vertex reordering in two key optimizations: degree field elimination and mini edge list optimization. The degree field elimination technique specifically targets vertices whose degrees exceed a configurable threshold $\degt$ (classified as large vertices), while the mini edge list optimization handles only those with degrees at or below the threshold (designated as mini vertices). We use $v'_i$ to indicate the $i$-th vertex after reordering, and $id_v$ to represent the $id$ of vertex $v$. An SSD-stored \texttt{v2id} table is hence generated to record the corresponding relation. This table does not need to reside in memory and is only useful during program initialization and termination. ACGraph operates on the reordered graph.

\header{\bf Degree Field Elimination.} 
ACGraph removes the 4-byte degree overhead through virtual vertex insertion and vertex reordering. First, virtual vertices are created with their offsets pointing to fragmentation boundaries in blocks. Then, those virtual vertices and large vertices are sorted together by offset and reassigned a new $id$ based on their sequence.  For example, we generate a virtual vertex in \figref{fig:storage} (filled with gray color), and set its offset to indicate the end of the last vertex in block 1. This virtual vertex gets a $id$ of $i+1$ after reordering. The virtual vertices are distinguished by setting their offset's highest bit, and ACGraph provides \texttt{is\_virtual(v)} method to filter them  during traversal. In this way, original large vertices retain the $\mathtt{deg}(v'_i) = \mathtt{offset}(v'_{i+1}) - \mathtt{offset}(v'_i)$ invariant, while virtual vertices keep unreached to preserve algorithm correctness, as they are not visited. Because the number of blocks is typically much smaller than the number of vertices, the additional space introduced is far lower than that of the degree field.

\header{\bf Mini Edge List Optimization.} 
\revise{}
It can be noticed that edge lists no more than 8-byte can be embedded directly within the index field of corresponding vertices, bringing no extra memory overhead, as what \cite{chunkgraph} does. 
However, this approach cannot be directly adapted to ACGraph, as the absence of the degree field (which is needed in \cite{chunkgraph}) prevents distinguishing whether the offset field encodes a true offset or an edge list. We address this challenge and extend the optimization through the just-mentioned vertex reordering technique. To be specific, these mini vertices are sorted in descending degree order and assigned consecutive $id$s that directly follow those of large vertices. Their edge lists are compactly arranged in a dedicated $\mathtt{mini\_data}$ array, mirroring the pattern of the CSR format.

Then, to trace the degree and the offset of mini vertices, we construct an $\idt$ array of size $\degt+1$:
\begin{equation}
    \label{eq:idbound}
    \idt[deg] = \min_{i}\{ i \mid \mathtt{deg}(v'_i) \leq deg\}.
\end{equation}

The degree and edge list offset of a  mini vertex $v'_i$ can be computed rather than explicit storage. The degree calculation employs:
\begin{equation*}
    \mathtt{deg}(v'_i) = \max_{0 \leq deg \leq \degt} \{deg \mid \idt[deg] \leq i \},
\end{equation*}

While the offset in $\mathtt{data\_{mini}}$ is computed via:
\begin{equation*}
    \begin{aligned}
        \mathtt{offset}(v'_i) = &(i - \idt[\mathtt{deg}(v'_i) ]) \cdot \mathtt{deg}(v'_i)  + \\
        &\sum_{i=deg(v'_i) +1}^{\degt} (\idt[i-1] - \idt[i]) \cdot i,
    \end{aligned}
\end{equation*}
Next, we include an example to illustrate how it works. 
\begin{example}
    Assume $\degt=3$ and there are 10 large vertices, 500 vertices of degree 3, 1000 of degree 2, and 2000 of degree 1. Then, $\idt[3]=10, \idt[2]=510, \idt[1]=1510, \idt[0]=3510$.
    Consider a vertex $v'_{1200}$, i.e., the reordered ID is 1200. After checking the $\idt[2]$, it is the maximum degree such that $\idt[2]\leq 1200$. Hence, the degree of $v'_{1200}$ is 2. 
Then, to derive the offset in the $\mathtt{data\_{mini}}$ array, it is computed as $(510-10)\times 3+(1200-510)\times 2$.
\end{example}

Normally, the degree threshold $\degt$ is a constant value of $2$ or $3$. This architecture eliminates per-node degree storage overhead while maintaining $O(1)$ access complexity through algebraic computation. The $\degt$  enables explicit trade-off between memory consumption and I/O overhead.

\section{Experiments}
\label{sec:exp}
In this section, we compare our solution {\proposed} against two state-of-the-art (SOTA) out-of-core graph processing systems, Blaze \cite{Blaze} and CAVE \cite{CAVE}. \reviseRThree{We further compare against Galois \cite{galois}, a popular in-memory asynchronous system, which highlights the performance-scalability trade-offs across in-memory and out-of-core approaches}. We implement our {\proposed} in C++ \cite{code_url}. For a fair comparison, we use the open-source implementations of Blaze\footnote{\url{https://github.com/NVSL/blaze}}, CAVE\footnote{\url{https://github.com/BU-DiSC/CAVE}} \reviseRThree{and Galois\footnote{\reviseRThree{\url{https://github.com/IntelligentSoftwareSystems/Galois}}}}. All code is compiled by {g++ 11.5.0} with the {-O3} flag. \reviseROne{Additionally, the \texttt{O\_DIRECT} flag is enabled to bypass the OS page cache, ensuring data is read directly from disk to userspace.}

\subsection{Experimental Settings}
\textbf{Graph Algorithms.} \revise{We evaluate our system using five popular graph algorithms that admit asynchronous execution: {breadth first search (BFS)}, {weakly connected component (WCC)}, {$k$-core}, {single source personalized PageRank (SSPPR)}, and PageRank (PR)}. For WCC, we adopt \textit{label propagation (LP)} \cite{10.1145/3159652.3159696} algorithm, which is a common algorithm used by graph processing systems \cite{Blaze}. For SSPPR, we use the {forward push (FP)} algorithm \cite{andersen2007local, HouCWW21, HouGZ0W23}, with probability $\alpha = 0.15$ and threshold $r_{\max} = 10^{-9}$. As a special case of personalized PageRank, PageRank set the initial distribution to $\{\frac{1}{n}, \frac{1}{n}, \cdots, \frac{1}{n}\}$, and then employs the same algorithm as SSPPR with $r_{\max} = 10^{-10}$. The parameter $k$ of $k$-core algorithm is set to $10$. To ensure a fair comparison, all systems use the same source vertices when running  BFS and SSPPR queries. The cache pool size for {\proposed}, CAVE, and Blaze is set to 256 MB by default. However, we observe that {\proposed} becomes insensitive to cache pool size once it exceeds a modest threshold (e.g., 40 MB). For fairness, we maintain the same cache size across all methods in our comparisons.

\header
\textbf{Hardwares.} All experiments are conducted on a Linux server running Ubuntu 20.04.6 LTS with kernel version 5.8.0. The system is equipped with 377GB of main memory and two {Intel(R) Xeon(R) Gold 5218 CPUs @ 2.30GHz}, each featuring 16 physical cores (32 threads with hyper-threading). The datasets are stored on a 1TB PCIe SSD device. To ensure consistency, we run each experiment three times, and report the average of the results \cite{CAVE}.

\header
\textbf{Datasets.} We evaluate our solution using four real-world graphs: Twitter, Friendster, UK-Union and ClueWeb12. These datasets are widely adopted in benchmarking graph processing frameworks, as seen in prior works \cite{GridGraph, corograph, ligra, Blaze, clip, DBLP:conf/kdd/ParkMK16, DBLP:journals/pvldb/GillDHPP20}. Twitter and Friendster are social networks, while the others are web graphs. Their key characteristics are summarized in \tabref{tab:dataset}. The graph sizes are reported based on the \textit{CSR (compressed sparse row)} format, where the offset array uses 8-byte unsigned integers and the edges array uses 4-byte unsigned integers. In the type column, `U' denotes an undirected graph, and `D' denotes a directed graph. To meet the requirement of algorithms that require undirected graph inputs (e.g., WCC and $k$-core), we preprocess the graphs accordingly by generating symmetrized edges. CAVE cannot process UK-Union and ClueWeb12 due to its use of 4-byte integers for offsets in the open-source implementation, so results on these datasets are omitted. \revise{In \tabref{tab:dataset}, $\boldsymbol{T}_p$ denotes the partitioning and reordering time of {\proposed} for preprocessing, measured in \emph{seconds}. On the largest dataset, ClueWeb12 (43B edges), it takes only 166 seconds.}

\begin{table}[!t]
\caption{Summary of the tested datasets.}\label{tab:dataset}
\vspace{-3mm}
\scalebox{0.95}{
\begin{tabular}{@{}lcccccc@{}}
\toprule
\textbf{Dataset} & \textbf{Abbr.} & $\boldsymbol{|V|}$ & $\boldsymbol{|E|}$ & \textbf{Size} & \textbf{Type} & $\mathbf{T_{p}}$   \\ \midrule
Twitter \cite{kwak2010twitter} & TW & 42M & 1.5B & 5.8GB & D & 6.2 \\
Friendster \cite{friendster} & FR & 66M & 3.6B & 15GB & U & 12.4  \\
UK-Union \cite{ukunion} & UK & 133M & 5.5B & 22GB & D & 21.4  \\
ClueWeb12 \cite{BMSB} & CL & 978M & 43B & 166GB & D & 166.2  \\ \bottomrule
\end{tabular}
}
\vspace{-3mm}
\end{table}

\header
\textbf{Metrics.} We use a variety of metrics to measure the performance of the framework, including:
\begin{itemize}[leftmargin=*]
    \item \textit{End-to-End Runtime}. Total execution time is measured from program launch to algorithm completion, including system initialization, graph loading and building, vertex state initialization, initial frontier construction, and algorithm execution phases. \reviseRThree{For Galois, we report only the algorithm execution time, as it is an in-memory system and places less emphasis on graph loading.}
    \item \textit{Memory Footprint}. Peak memory consumption is measured by the Maximum Resident Set Size (max RSS) metric, which is obtained from the \texttt{/bin/time -v -p} command.
    \item \textit{Disk I/O.} Disk read operations are quantified as $\Delta_{\text{I/O}} = \texttt{iostat}_{\text{end}} - \texttt{iostat}_{\text{start}}$, where $\texttt{iostat}_{\text{end}}$ and $\texttt{iostat}_{\text{start}}$ represent the pre-execution and post-execution measurements of cumulative disk I/O operations, respectively recorded via the \texttt{iostat} utility.
\end{itemize}

\begin{figure*}[t]
	\centering
		\begin{tabular}{c@{\hspace{-3mm}}c@{\hspace{-3mm}}c@{\hspace{-3mm}}c@{\hspace{-3mm}}c}
             \multicolumn{5}{c}{\includegraphics[height=3mm]{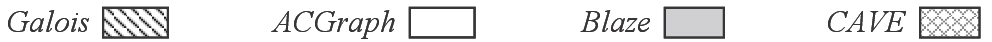}}  \\[-1mm]
			\includegraphics[width=30mm]{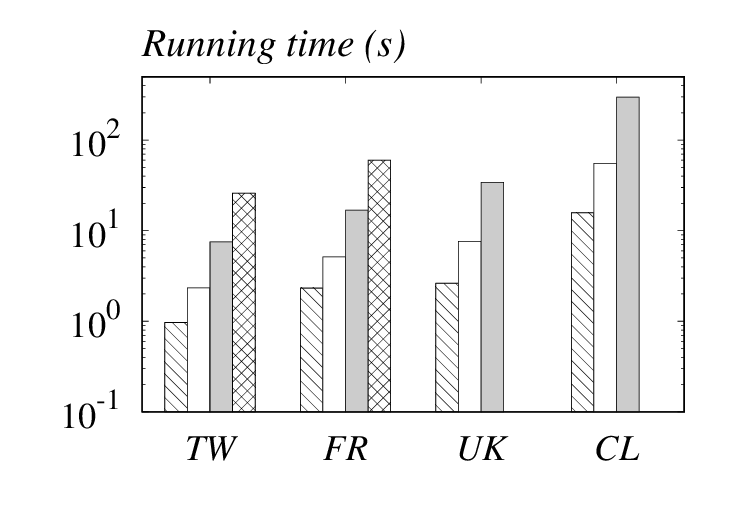} &
			\includegraphics[width=30mm]{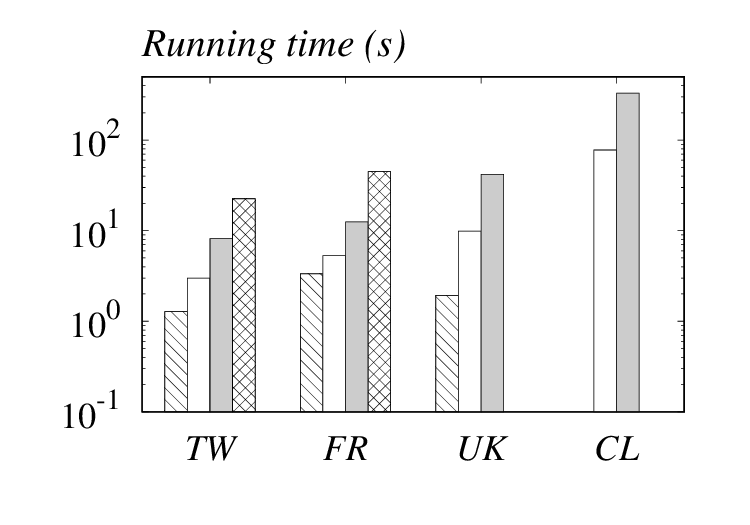} &
			\includegraphics[width=30mm]{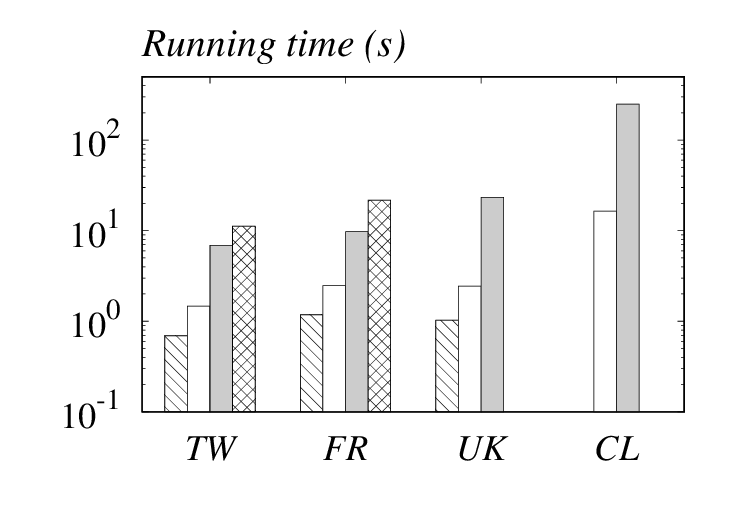} &
			\includegraphics[width=30mm]{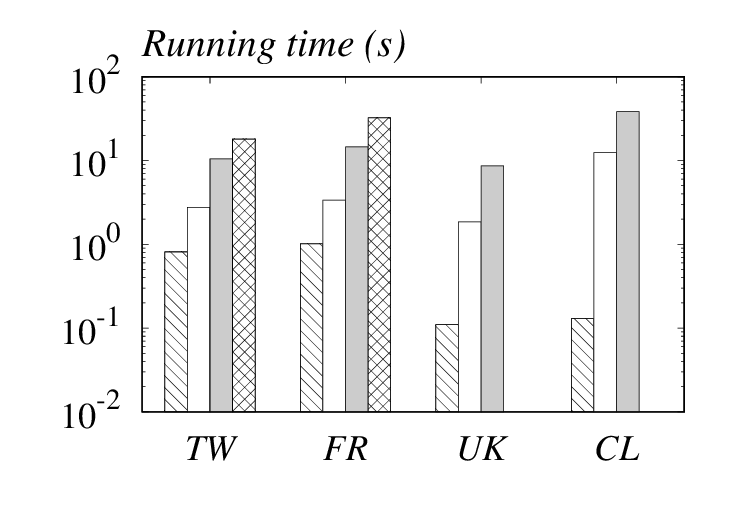} &
            \includegraphics[width=30mm]{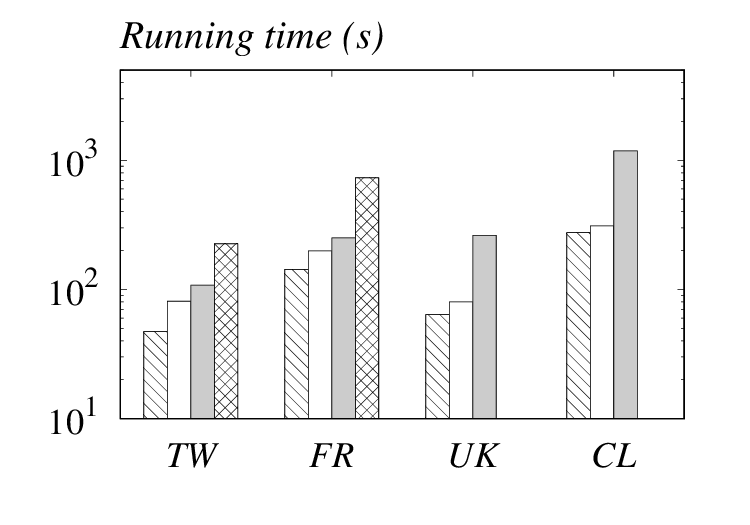} 
			\\[-2mm]
            (a) BFS &
			(b) WCC &
			(c) $k$-core &
			(d) SSPPR  &
            (e) PageRank
            \\[-1.5mm]
		\end{tabular}
		\vspace{-2mm}
		\caption{Running time of each graph processing system (lower is better).} \label{fig:runningtime}
		\vspace{-3mm}
\end{figure*}

\begin{figure*}[t]
	\centering
		\begin{tabular}{c@{\hspace{-3mm}}c@{\hspace{-3mm}}c@{\hspace{-3mm}}c@{\hspace{-3mm}}c}
			\includegraphics[width=30mm]{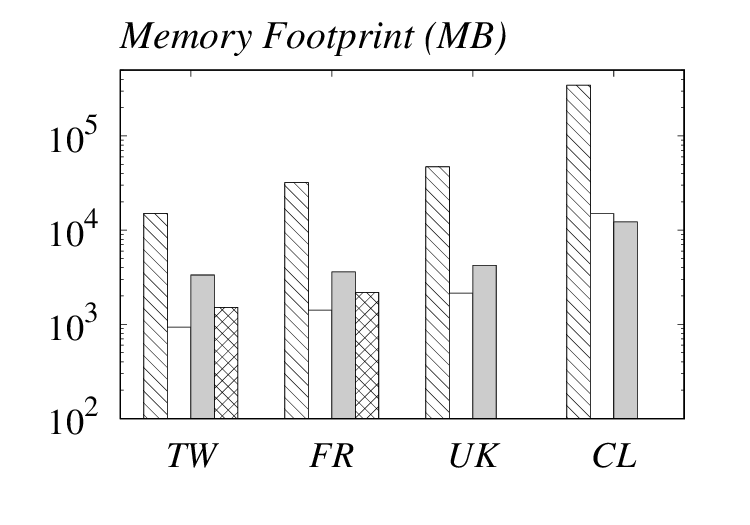} &
			\includegraphics[width=30mm]{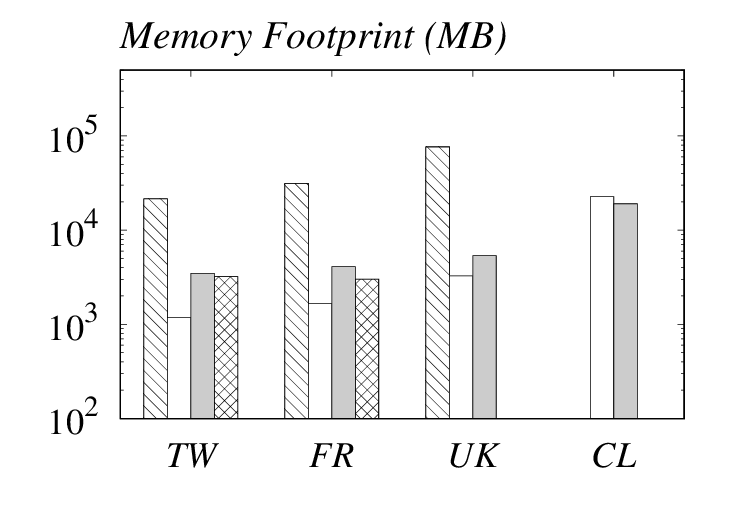} &
			\includegraphics[width=30mm]{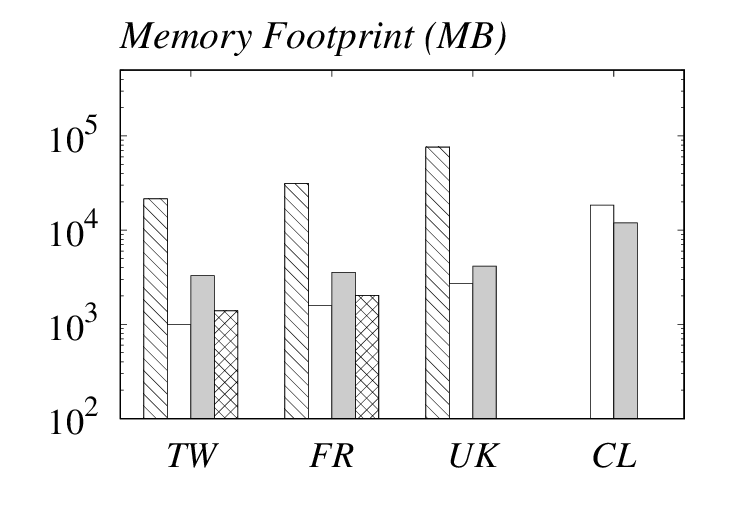} &
			\includegraphics[width=30mm]{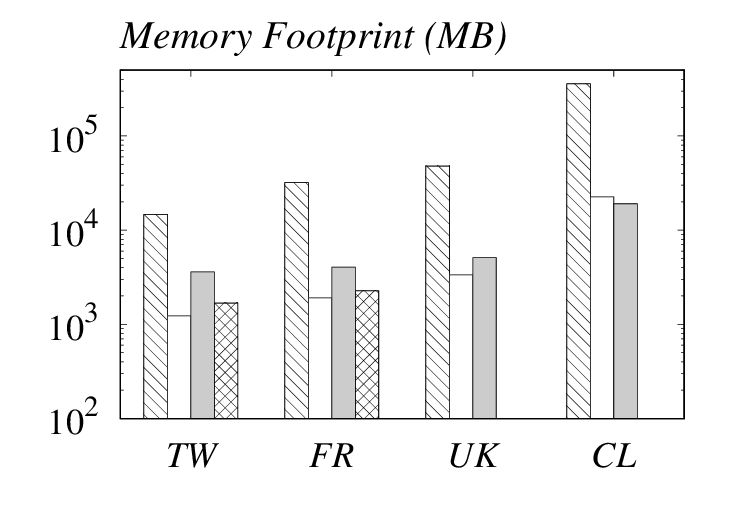} &
            \includegraphics[width=30mm]{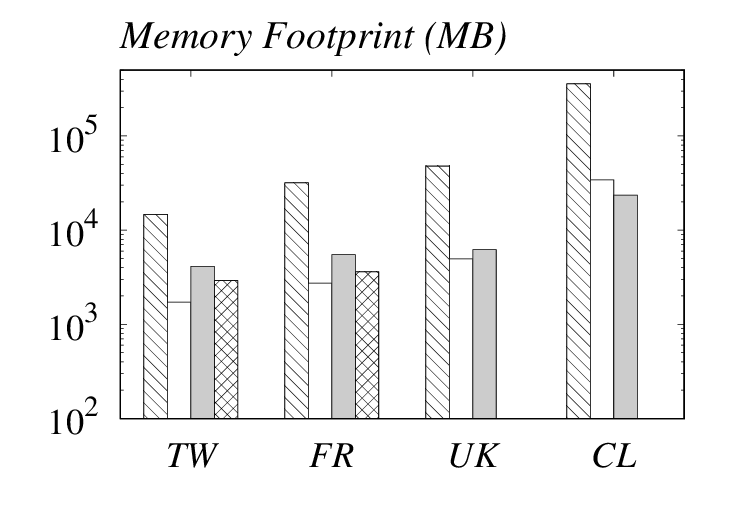}
			\\[-2mm]
            (a) BFS &
			(b) WCC &
			(c) $k$-core &
			(d) SSPPR  & 
            (e) PageRank
            \\[-1.5mm]
		\end{tabular}
		\vspace{-2mm}
		\caption{Memory footprint of each graph processing systems. } \label{fig:memory}
		\vspace{-3mm}
\end{figure*}

\subsection{Overall Performance} \label{sec:overall}
First, we test \reviseROne{five graph algorithms (i.e., BFS, WCC, $k$-core,  SSPPR and PR) that support asynchronous execution}, to evaluate the performance of each GPS in terms of running time, as shown in \figref{fig:runningtime}. The reported running time of out-of-core GPSs includes initialization time (e.g., index loading), while the in-memory GPS does not include this time. As shown in \figref{fig:runningtime}, {\proposed} consistently outperforms Blaze and CAVE across all tested datasets and algorithms. In particular, {\proposed} achieves notable speedups over the second-fastest out-of-core system, Blaze: \reviseROne{up to 5.50$\times$ (BFS), 4.26$\times$ (WCC), 15.21$\times$ ($k$-core), 4.67$\times$ (SSPPR) and 3.80$\times$ (PR).} Compared to CAVE, {\proposed} yields \reviseROne{up to 12.34$\times$, 10.75$\times$, 8.80$\times$, 9.62$\times$ and 3.68$\times$ improvement , respectively}. \reviseROne{Moreover, across the four algorithms excluding locally-focused SSPPR, {\proposed} achieves higher speedup ratios on web graphs than on social networks. This stems from web graphs' larger diameters, which require more iterations to coverage and incur greater synchronization overhead.} These significant improvements can be attributed to our asynchronous design, which effectively addresses the limitations discussed in Sec.~\ref{sec:motivation}. \reviseRThree{Galois consistently outperforms other frameworks: When executing BFS, WCC, $k$-core, SSPPR and PR, Galois achieves speedups of 2.20–3.51$\times$, 1.42–2.45$\times$, 2.10–2.38$\times$, 3.41-95.72$\times$ and 1.13–1.72$\times$ over {\proposed} respectively, benefiting from no I/O overhead. However, Galois is unable to run WCC and $k$-core on ClueWeb12, as these algorithms require undirected graph processing, which exceeds the machine’s memory limits. The limitation highlights the low scalability of in-memory systems.}

We also report the memory footprint of our {\proposed} against its three competitors, as illustrated in \figref{fig:memory}. We observe that {\proposed} consistently requires lower memory usage than others across all evaluated graph algorithms on the Twitter, Friendster, and UK-Union datasets. To explain, Blaze suffers from higher inherent memory overheads due to its implementation on top of the Galois framework \cite{galois}, while CAVE maintains the entire frontier set in array format without bitmap optimization. On ClueWeb12, ACGraph consumes slightly more memory than Blaze, mainly because it maintains a 64-byte metadata for each block and allocates an 8-byte offset field for every vertex. Since ClueWeb12 contains over 978 million vertices, this leads to additional memory overhead, as observed. Considering its superior performance of achieving up to 15$\times$ speedup over Blaze, it demonstrates a favorable trade-off between computational efficiency and memory usage. \reviseRThree{Compared to Galois, {\proposed} achieves a remarkable trade-off between performance and scalability. While incurring a performance penalty of 1–2$\times$ in most scenarios, it reduces memory cost by 7.49–27.95$\times$.}

\subsection{I/O Efficiency}

\begin{figure}[t]
	\centering
	\begin{small}
		\begin{tabular}{cc}
        \multicolumn{2}{c}{\hspace{-0mm}\includegraphics[height=2.6mm]{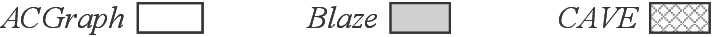}}  \\[-1mm]
        	\hspace{-3mm} \includegraphics[width=38mm]{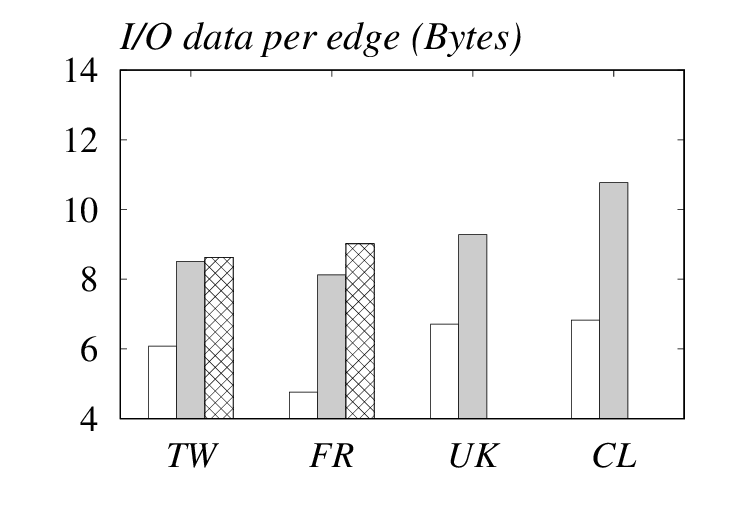} &
			\hspace{-3mm} \includegraphics[width=38mm]{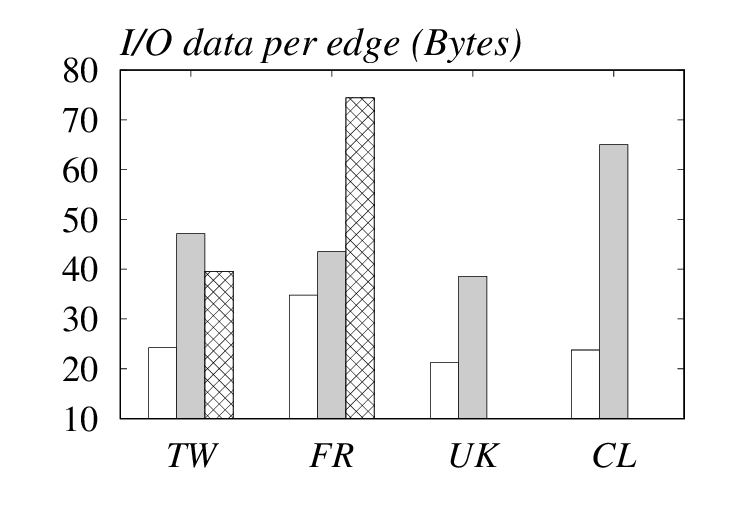} \\[-2mm]
			\hspace{-3mm} (a) BFS  &
			\hspace{-3mm} (b) SSPPR \\
		\end{tabular}
        \vspace{-4mm}
		\caption{ Read inflation (lower is better).} 
        \label{fig:readamp}
	\end{small}
    \vspace{-4mm}
\end{figure}

\begin{figure}[t]
	\centering
	\begin{small}
		\begin{tabular}{cc}
        	\hspace{-3mm} \includegraphics[width=38mm]{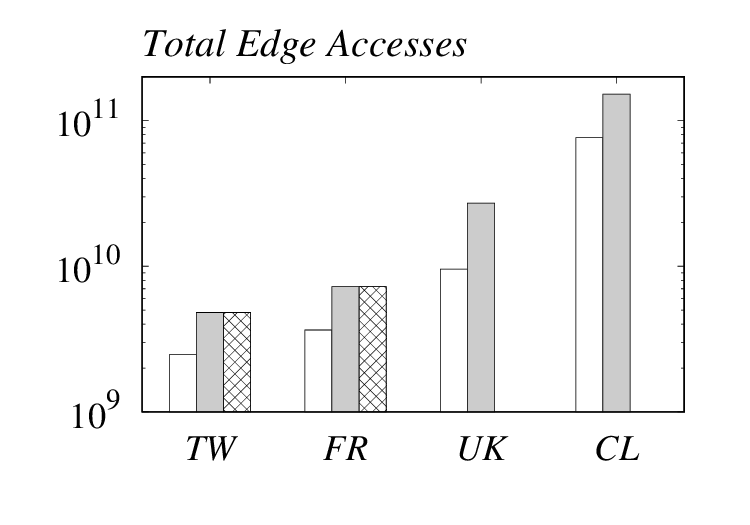} &
			\hspace{-3mm} \includegraphics[width=38mm]{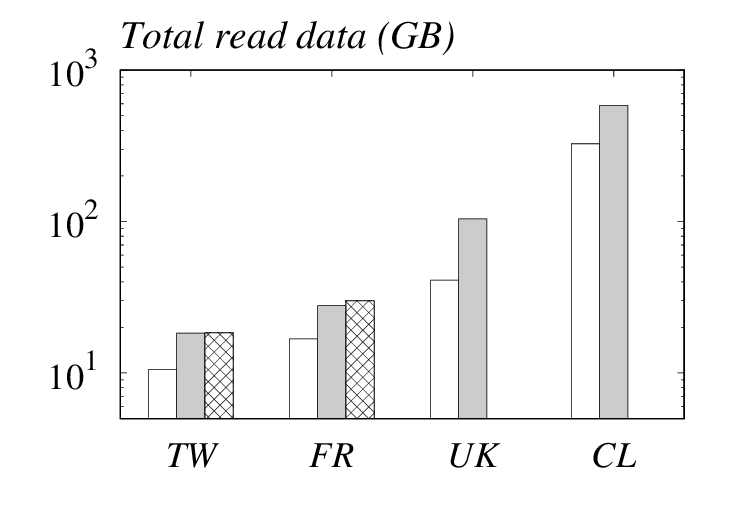} \\[-2mm]
			\hspace{-3mm} (a) Edge accesses &
			\hspace{-3mm} (b) I/O read volume \\
		\end{tabular}
        \vspace{-4mm}
		\caption{Work inflation (lower is better).} 
        \label{fig:workamp}
	\end{small}
    \vspace{-4mm}
\end{figure}

\begin{figure}[t]
	\centering
	\begin{small}
		\begin{tabular}{cc}
        	\hspace{-3mm} \includegraphics[width=38mm]{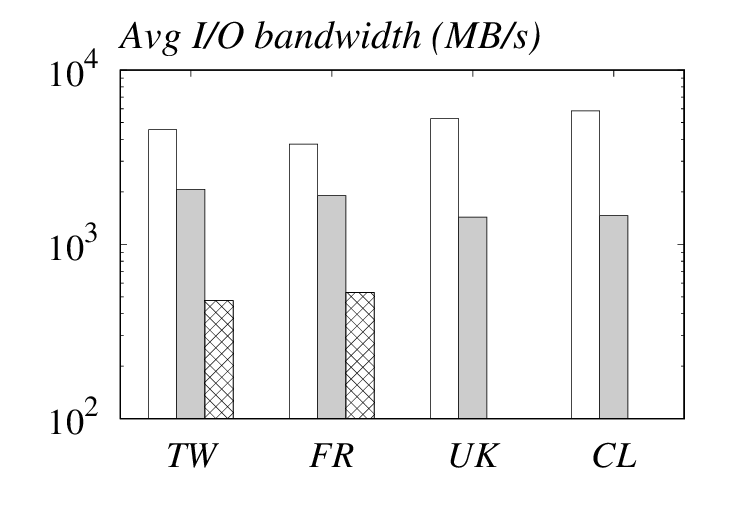} &
			\hspace{-3mm} \includegraphics[width=38mm]{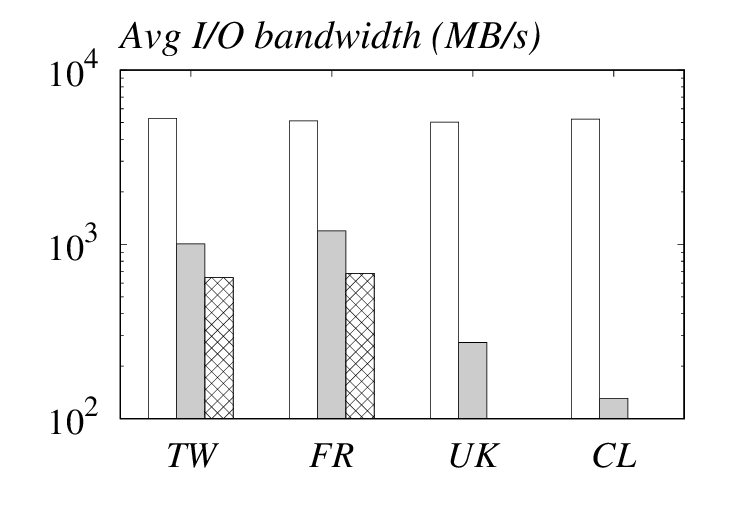} \\[-2mm]
			\hspace{-3mm} (a) BFS  &
			\hspace{-3mm} (b) $k$-core \\
		\end{tabular}
        \vspace{-4mm}
		\caption{I/O Throughput (higher is better).} 
        \label{fig:avgtp}
	\end{small}
    \vspace{-4mm}
\end{figure}

\textbf{Read Inflation.}
We further examine how {\proposed} addresses read inflation issues during BFS and SSPPR execution compared to the two out-of-core GPSs: Blaze and CAVE. To quantify the extent of read inflation, we measure the average I/O volume per edge access. Note that each edge is represented using 4 bytes. Since each edge is accessed exactly once in BFS, the theoretical minimum I/O volume per edge is 4 bytes. The experimental results in \figref{fig:readamp} show that {\proposed} achieves superior I/O efficiency, requiring fewer than 7 bytes per edge on average. On Friendster, this improves to just 4.8 bytes per edge, approaching the theoretical minimum. In contrast, Blaze incurs at least 8.1 bytes per edge across all four datasets, while CAVE requires a minimum of 8.6 bytes per edge. For SSPPR, all frameworks experience increased read inflation. Nevertheless, {\proposed} maintains a clear advantage, using only $37\%$ of Blaze’s I/O volume per edge on ClueWeb12.

\header \textbf{Work Inflation.} \figref{fig:workamp} illustrates the work inflation issue in WCC algorithms. As discussed in \secref{sec:motivation}, the synchronous semantics of Blaze and CAVE lead to near-full graph loading in early iterations. In contrast, {ACGraph} converges faster with significantly fewer edge accesses by leveraging priority-based scheduling. On Twitter and Friendster, CAVE and Blaze process 1.95$\times$ and 1.99$\times$ more edges, respectively. For UK-Union and ClueWeb12, Blaze incurs $2.84$x and $1.98$x more edge accesses than {ACGraph}. Hence, {ACGraph} reduces I/O access volume by up to $60\%$ compared to Blaze and $44\%$ compared to CAVE, confirming its effectiveness in mitigating work inflation.

\header \textbf{I/O Throughput.} \figref{fig:avgtp} shows the average bandwidth during $k$-core and BFS execution. ACGraph achieves a bandwidth of 4.5~GB/s and 3.7~GB/s for BFS on Twitter and Friendster respectively and over 5.0~GB/s in all other cases. These bandwidths approach the maximum 6.0~GB/s sequential access speed of our SSD hardware, confirming that ACGraph effectively saturates the storage bandwidth. In comparison, CAVE is limited to no more than 800~MB/s across all test cases due to inefficient single-threaded task collection and distribution. While Blaze achieves over 2.0~GB/s for BFS and over 1.0~GB/s for $k$-core on Twitter and Friendster, its performance degrades significantly on UK-Union and ClueWeb12, due to longer convergence on these two datasets. In particular, Blaze requires 12 -- 24 iterations for BFS and $k$-core on Twitter and Friendster, while it takes 290 -- 670 iterations on UK-Union and ClueWeb12.

\revise{ 
\subsection{Synchronous Execution}

\begin{figure}[t]
	\centering
	\begin{small}
		\begin{tabular}{cc}
        \multicolumn{2}{c}{\hspace{-2mm}\includegraphics[height=2.8mm]{Figure/gps_legend_rev.eps}}  \\[-1mm]
        	\hspace{-3mm} \includegraphics[width=38mm]{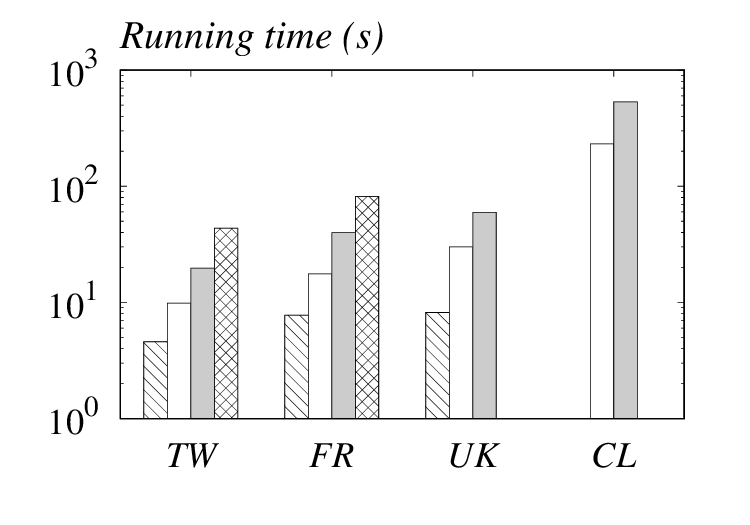} &
			\hspace{-3mm} \includegraphics[width=38mm]{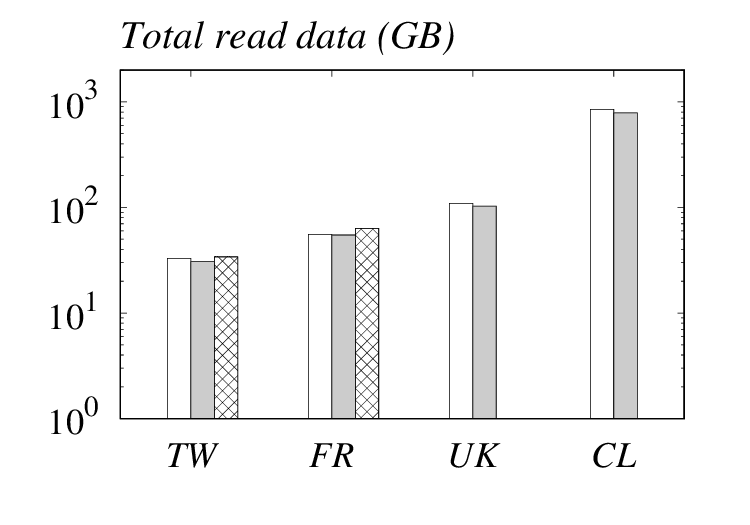} \\[-1mm]
			\hspace{-3mm} (a) Running Time  &
			\hspace{-3mm} (b) I/O Read Volume \\
		\end{tabular}
        \vspace{-3mm}
		\caption{\reviseROne{Synchronous execution: MIS.}}
        \label{fig:mis}
	\end{small}
    \vspace{-3mm}
\end{figure}

As discussed in \secref{sec:sync}, some algorithms inherently require synchronous execution. To evaluate {\proposed}'s performance in this context, we use the Maximal Independent Set (MIS) problem as a case study. Specifically, we adopt Blelloch’s Alg.\ 2 \cite{MIS}, ensuring determinism and comparability by fixing random seeds across all implementations. For this experiment, {\proposed} is configured to use its synchronous execution interface. Note that Galois fails to execute the algorithm on the ClueWeb12 dataset due to the memory requirements exceeding our machine’s capacity.

As shown in Fig.~\ref{fig:mis}, across all graphs, the out-of-core systems show similar I/O volumes (the difference is within $16\%$), as they access the same number of edges. Thanks to its asynchronous I/O pipeline, which boosts edge loading and processing throughput, {\proposed} achieved 2.0–2.3$\times$ speedup over Blaze. However, in synchronous mode, {\proposed} is also subject to synchronization stalls. As a result, the runtime ratios among the out-of-core systems remain similar across all four datasets. This contrasts with the results in \secref{sec:overall}, where {\proposed} achieves larger speedups on large-diameter web graphs, highlighting the  benefits of asynchronous execution.
}

\subsection{Tuning Parameters}
\begin{figure}[t]
	\centering
	\begin{small}
		\begin{tabular}{cc}
        \multicolumn{2}{c}{\hspace{-0mm}\includegraphics[height=2.8mm]{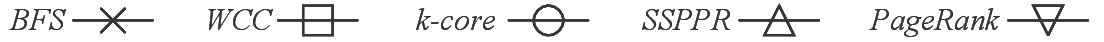}}  \\[-1mm]
        	\hspace{-3mm} \includegraphics[width=38mm]{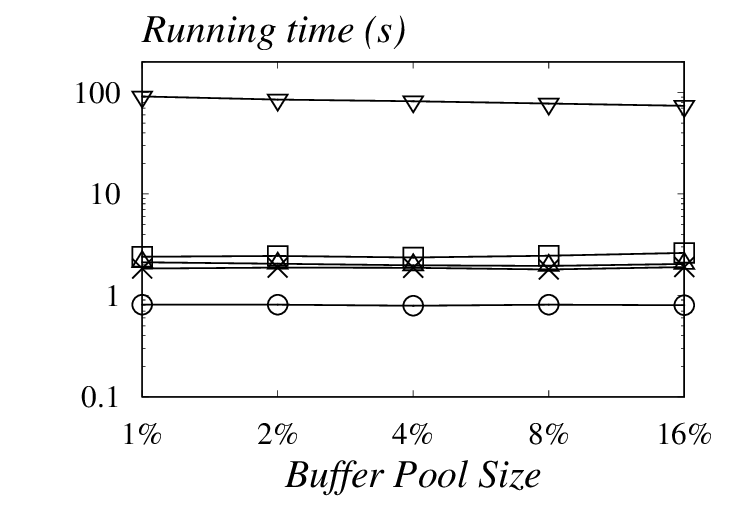} &
			\hspace{-3mm} \includegraphics[width=38mm]{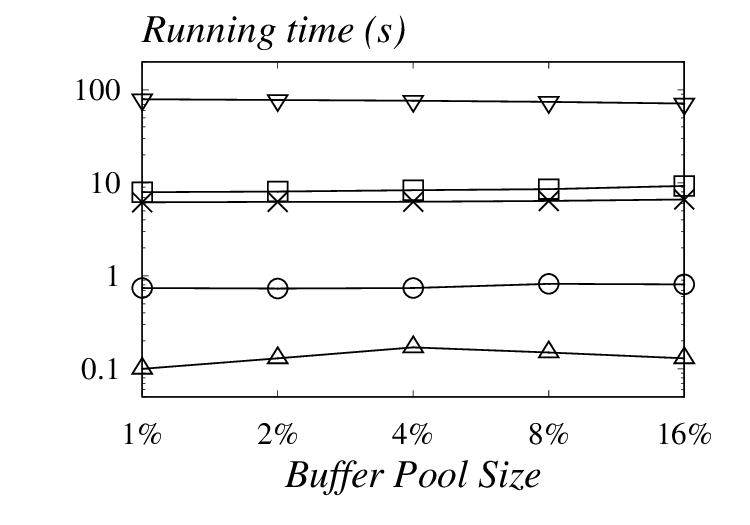} \\[-1mm]
			\hspace{-3mm} (a) Twitter  &
			\hspace{-3mm} (b) UK-Union \\
		\end{tabular}
        \vspace{-3mm}
		\caption{Performance sensitivity to buffer pool size.} 
        \label{fig:cache}
	\end{small}
    \vspace{-3mm}
\end{figure}

\begin{figure}[t]
	\centering
	\begin{small}
		\begin{tabular}{cc}
        	\hspace{-2.7mm} \includegraphics[width=38mm]{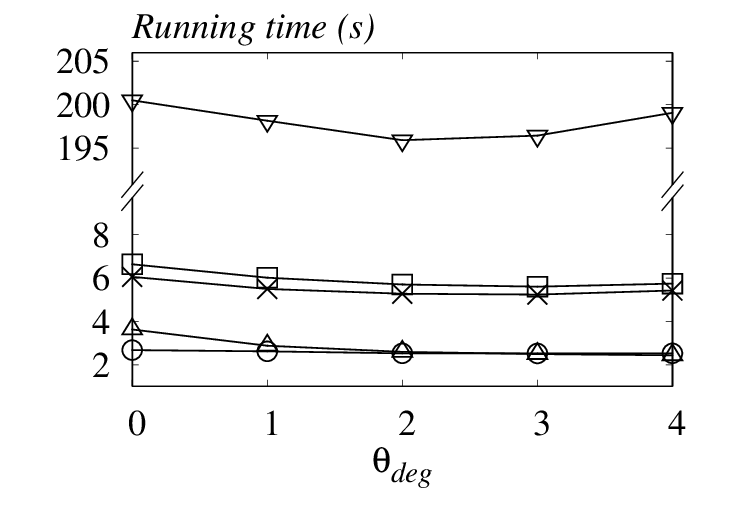} &
			\hspace{-2.7mm} \includegraphics[width=38mm]{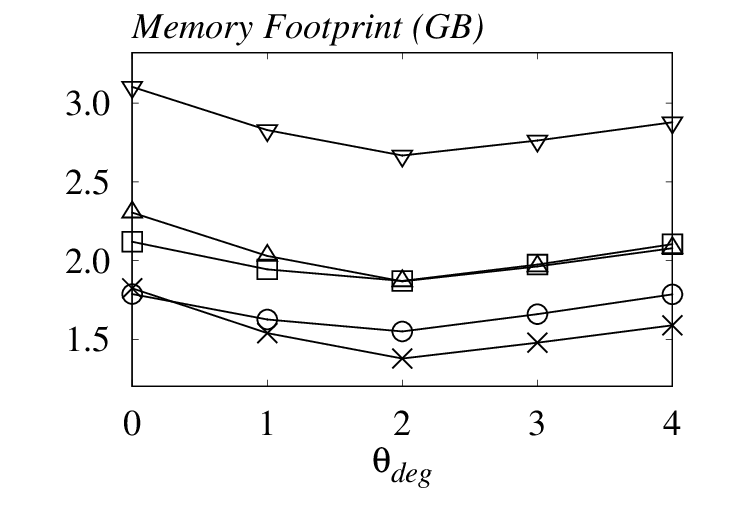} \\[-2mm]
			\hspace{-2.7mm} (a) Running Time  &
			\hspace{-2.7mm} (b) Memory Footprint \\
		\end{tabular}
        \vspace{-3mm}
        \caption{Perf. sensitivity to degree threshold on FR.} 
        \label{fig:degbound}
	\end{small}
    \vspace{-3mm}
\end{figure}

\begin{figure*}[t]
	\centering
    \begin{tabular}{c@{\hspace{-3mm}}c@{\hspace{-3mm}}c@{\hspace{-3mm}}c@{\hspace{-3mm}}c}
             \multicolumn{5}{c}{\includegraphics[height=2.6mm]{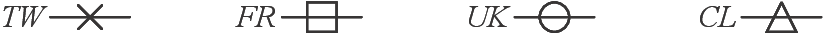}}  \\[-1mm]
        \includegraphics[width=30mm]{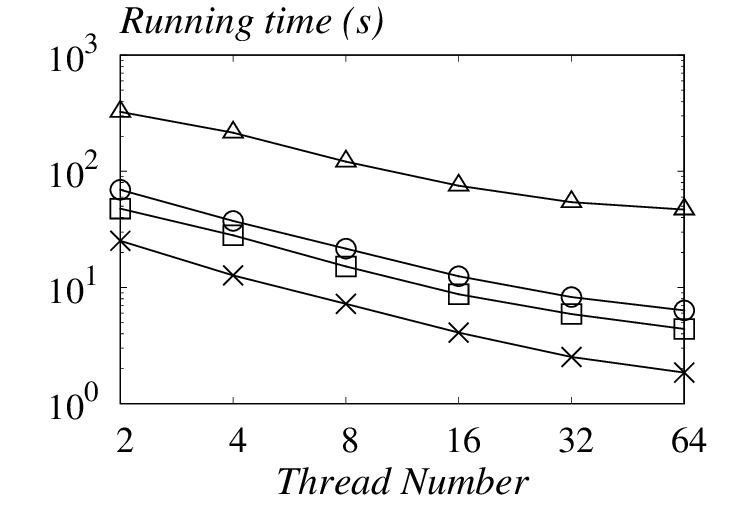} &
        \includegraphics[width=30mm]{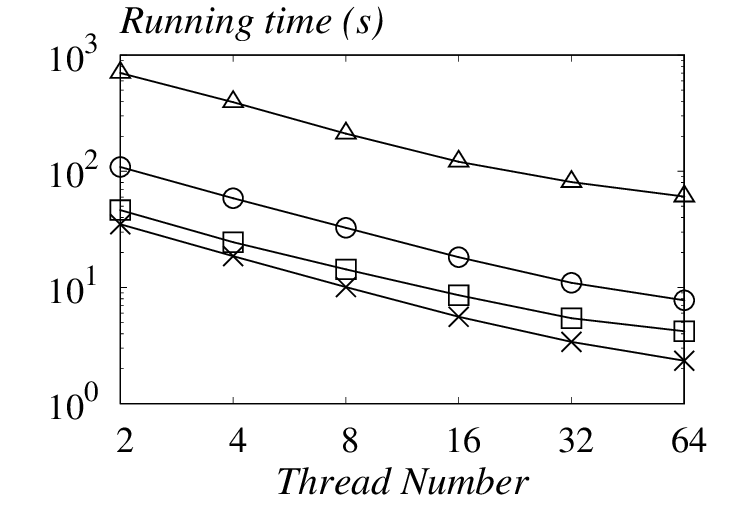} &
        \includegraphics[width=30mm]{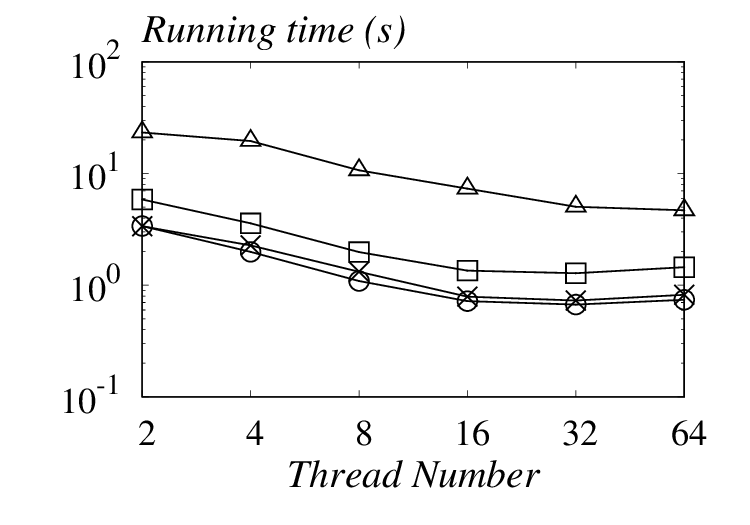} &
        \includegraphics[width=30mm]{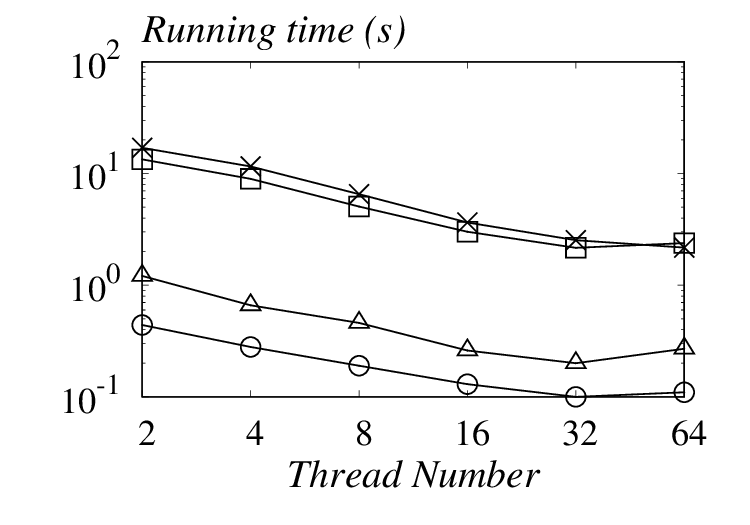} &
        \includegraphics[width=30mm]{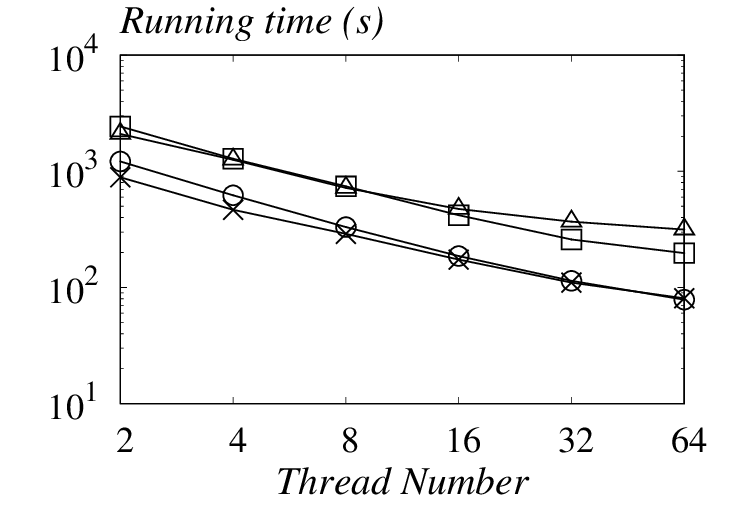}
        \\[-1mm]
        (a) BFS &
        (b) WCC &
        (c) $k$-core &
        (d) SSPPR &
        (e) PageRank \\[-1mm]
    \end{tabular}
    \vspace{-3mm}
    \caption{Algorithm execution time under varying numbers of threads.} \label{fig:thread_scaling}
    \vspace{-4mm}
\end{figure*}

\textbf{Buffer Pool Size.} We further examine the impact of the buffer pool size on the performance of {\proposed}, by varying the pool size from $1\%$ to $16\%$ of the input graph size, as shown in \figref{fig:cache}. Our results demonstrate that {\proposed} delivers stable and efficient performance across all configurations. This robustness stems from its effective graph locality preservation and efficient block reuse mechanism, which enable excellent performance even with minimal buffer allocation. Besides, we observe a marginal runtime increase at larger pool sizes, which is due to the initialization and maintenance overhead associated with expanded pool size.

\header
\textbf{Configurable Degree Threshold.}
\figref{fig:degbound} shows the performance of {\proposed} in terms of runtime and memory cost on dataset Friendster, as a function of the degree threshold $\degt$. The default 64-byte metadata layout only supports the settings of $\degt \geq 2$. For $\degt=1$, we increase the metadata size to 128 bytes, and for $\degt = 0$, further to 196 bytes, ensuring cacheline alignment. Consequently, the space usage is minimized at $\degt = 2$. This is because for $\degt < 2$, excessive space is occupied by metadata, and for $\degt > 2$, the mini edge list dominates the space overhead. Furthermore, \figref{fig:degbound} shows that while the execution time generally decreases with larger $\degt$, the reduction becomes marginal when $\degt$ exceeds 2. Therefore, balancing time and space efficiency, we set $\degt = 2$ by default.

\header \textbf{Thread Scaling.} \figref{fig:thread_scaling} presents the execution time of our {\proposed} on four graph algorithms, with the number of worker threads varying from 2 to 64. Note that the time spent in the initialization phase, which is executed in a single thread, is excluded from the reported measurements. As shown, when the disk is not saturated, {\proposed} demonstrates a near-linear performance improvement as the number of threads increases, achieving up to a 14.9$\times$s speedup.

\reviseRTwo{
\subsection{Ablation Study on the Partitioner}

\begin{table}[t]
\centering
\begin{small}
\caption{\reviseRTwo{Perf. ratio (BF/LPLF) based on I/O read volume and running time, with 1 indicating equal performance.} }
\vspace{-3mm}
\begin{tabular}{lccccc}
\hline
 & \textbf{BFS} & \textbf{WCC} & \boldmath$k$\textbf{-core} & \textbf{SSPPR} & \textbf{PageRank} \\
\hline
UK-Union I/O   & 1.16 & 1.26 & 0.59 & 2.22 & 1.04 \\
UK-Union Time  & 1.14 & 1.26 & 0.62 & 1.75 & 1.03 \\
\hline
ClueWeb12 I/O   & 1.31 & 1.17 & 0.59 & 1.79 & 1.21 \\
ClueWeb12 Time & 1.28 & 1.32 & 0.73 & 1.75 & 1.02 \\
\hline
\end{tabular}
\label{tab:io}
\end{small}
\vspace{-2mm}
\end{table}

\begin{figure}[t]
	\centering
	\begin{small}
		\begin{tabular}{cc}
        \multicolumn{2}{c}{\hspace{-0mm}\includegraphics[height=2.7mm]{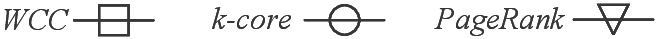}}\\[-1.5mm]
        	\hspace{-3mm} \includegraphics[width=37mm]{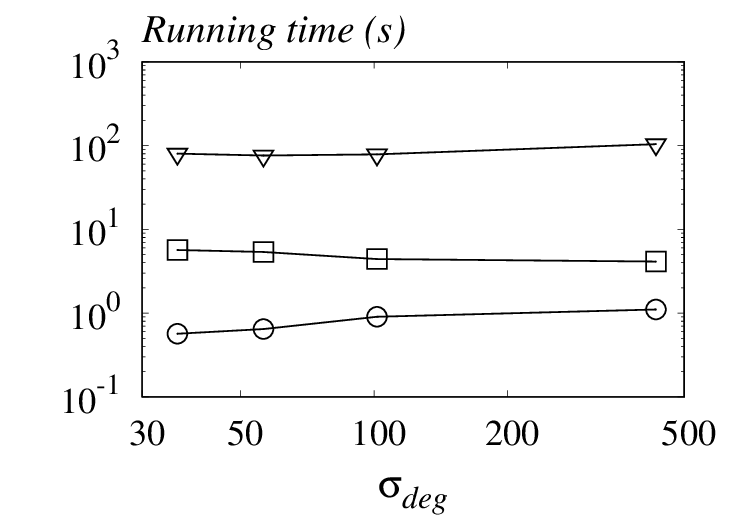} &
			\hspace{-3mm} \includegraphics[width=37mm]{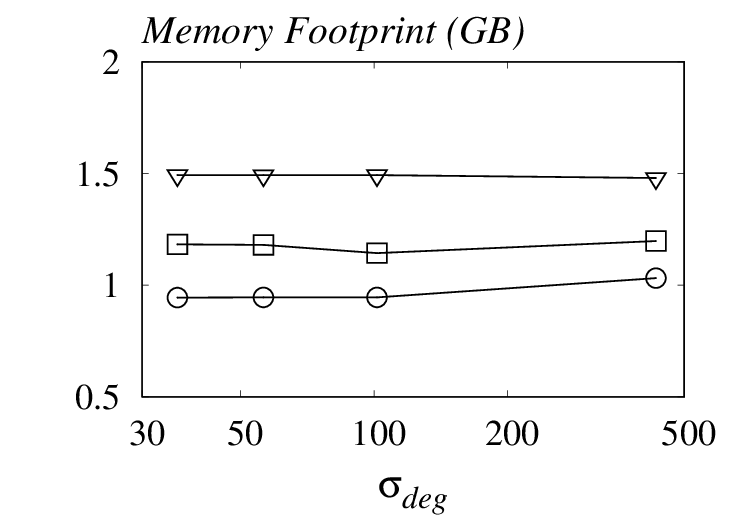} \\[-1.5mm]
			\hspace{-3mm} (a) Running Time  &
			\hspace{-3mm} (b) Memory Footprint \\
		\end{tabular}
        \vspace{-3mm}
        \caption{\reviseRTwo{Perf. sensitivity to degree skewness, where $\sigma_{deg}$ represents standard deviation of graph degrees.}} 
        \label{fig:degskew}
	\end{small}
    \vspace{-2mm}
\end{figure}

We conduct experiments to explain why we adopt the locality-preserving last-fit (LPLF) partitioner as the default. While partitioning is not the main focus of our framework, we seek a lightweight and robust solution. We show that LPLF is more I/O- and time-efficient than simpler alternatives such as the degree-sorted best-fit (BF) packing method, where vertices are processed in descending degree order and assigned to the tightest available block, creating new blocks when necessary. \tabref{tab:io} reports the ratio of disk reads and runtime for BF relative to LPLF on the two large datasets (i.e., UK-Union and ClueWeb12). In the table, values greater than 1 indicate that BF performs worse. LPLF outperforms BF in 4 out of 5 algorithms, showing better robustness. BF performs better only for $k$-core due to better alignment with the algorithm’s access pattern. For other algorithms, the loss of locality degrades performance. As general-purpose systems demand consistent performance across workloads, we adopt LPLF by default while allowing users to implement custom partitioners for specific use cases.

We further evaluate the impact of degree skewness on {\proposed} using four synthetic graphs (all with 64M vertices, 1.5B edges) generated via R-MAT~\cite{rmat}, with parameters adjusted to vary skewness. As shown in \figref{fig:degskew}, the standard deviation of vertex degrees ranges from 30 to 500, resulting in skewness scores of 2.12, 3.68, 8.97, and 79.77. We report query times only for global asynchronous algorithms, as PPR and BFS are highly sensitive to source vertices and thus less generalizable. {\proposed} shows stable performance across all graphs, underscoring  the robustness of its preprocessing and overall design. Finally, {\proposed} exhibits consistent preprocessing times ($\approx$10 seconds) across all graphs due to their identical sizes.

}

\section{Related Work} \label{sec:related}
Beyond the out-of-core GPSs discussed in \secref{sec:oocgps}, prior work falls into two main categories: in‑memory GPSs and distributed GPSs.

\noindent
{\bf In-memory GPSs.} Ligra \cite{ligra} introduces a lightweight vertex‐centric API for multicore machines, switching automatically between push and pull traversals to balance work and minimize overhead. GPOP \cite{gpop} adopts a partition‐centric model that partitions the graph into fine‐grained blocks, reducing synchronization and improving cache utilization. Corograph \cite{corograph} employs a hybrid execution model to achieve both cache efficiency and work efficiency for batched query processing, which differs from our setting. ForkGraph \cite{forkgraph} partitions the graph into cache‑sized blocks and processes query batches concurrently, boosting cache hit rates and overall throughput via carefully designed scheduling strategies. \reviseRTwo{GraphLab \cite{graphlab} pioneers an asynchronous, shared-memory, graph-parallel framework for iterative machine learning algorithms. }
\reviseRThree{Galois \cite{galois} is a popular in-memory asynchronous GPS that provides an optimistic execution engine with fine‐grained parallelism, priority scheduling and work stealing, allowing developers to express algorithms at a high level. The design philosophies of these in-memory asynchronous GPSs, such as the vertex-centric paradigm, have influenced subsequent GPSs. However, as noted in \secref{sec:intro}, they require substantial redesign to perform well in out-of-core environments.
}

\header{\bf Distributed GPSs.}
Distributed GraphLab \cite{distgraphlab} scales the GraphLab paradigm to clusters for efficient large graph processing on commodity clouds. Pregel \cite{pregel} popularized the Bulk Synchronous Parallel (BSP) vertex‐centric model via message passing and global barriers to process large graphs. PowerGraph \cite{powergraph} extends it with the Gather–Apply–Scatter (GAS) abstraction and vertex‐cut partitioning to handle power‐law graphs. GraphX \cite{graphx} builds on Spark RDDs to unify graph and dataflow operations with fault tolerance and query optimization. Gemini \cite{gemini} introduces multiple computation-centric optimizations to improve scalability. TurboGraph++ \cite{turbograph} employs three‑level parallelism and overlapping to maximize the CPU, disk, and network utilization in clusters.

Additionally, extensive work has explored graph processing on emerging hardware such as GPUs, with frameworks like BTS \cite{gts}, Lux \cite{lux}, Scaph \cite{scaph}, CGraph \cite{cgraph}, TGraph \cite{tgraph}, and Nezha \cite{nezha}. \revise{
Some out-of-core GPSs (e.g., \cite{Blaze,CAVE,chunkgraph}) introduce custom storage formats to reduce space usage, but as discussed in \secref{sec:storage}, these approaches have limitations, motivating our hybrid design.
}

\section{Conclusions}\label{sec:conclusion}
Motivated by the I/O inefficiencies in existing out-of-core GPSs, we propose ACGraph, a novel asynchronous GPS for SSDs built on our proposed block-centric execution model. It features a hybrid storage architecture optimized for space and computation. Extensive evaluation on real graphs with tens of billions of edges shows that ACGraph significantly outperforms existing SSD-based GPSs.

\begin{acks}
This work was supported by the Hong Kong RGC GRF grant (No. 14217322) and the 1+1+1 CUHK-CUHK(SZ)-GDSTC Joint Collaboration Fund (No. 2025A0505000045).
\end{acks}


\bibliographystyle{ACM-Reference-Format}
\bibliography{reference}

\appendix

\end{document}